\documentclass[journal]{vgtc}                % final (journal style)
\ifpdf%                                % if we use pdflatex
  \pdfoutput=1\relax                   % create PDFs from pdfLaTeX
  \usepackage{graphicx}                % allow us to embed graphics files
  \DeclareGraphicsExtensions{.pdf,.png,.jpg,.jpeg} % for pdflatex we expect .pdf, .png, or .jpg files
\else%                                 % else we use pure latex
  \usepackage{graphicx}                % allow us to embed graphics files
  \DeclareGraphicsExtensions{.eps}     % for pure latex we expect eps files
\fi%

\usepackage{microtype}                 % use micro-typography (slightly more compact, better to read)
\PassOptionsToPackage{warn}{textcomp}  % to address font issues with \textrightarrow
\usepackage{textcomp}                  % use better special symbols
\usepackage{mathptmx}                  % use matching math font
\usepackage{times}                     % we use Times as the main font
\usepackage{cite}                      % needed to automatically sort the references
\onlineid{1077}

\vgtccategory{Research}
\vgtcpapertype{application/design study}

\title{Visualization of Human Spine Biomechanics for Spinal Surgery}

\author{Pepe Eulzer, Sabine Bauer, Francis Kilian, Kai Lawonn}
\authorfooter{
\item
 Pepe Eulzer is with the University of Jena, Germany. E-mail: pepe.eulzer@uni-jena.de.
\item
 Sabine Bauer is with the University of Koblenz-Landau, Germany. E-mail: bauer@uni-koblenz.de.
\item
 Francis Kilian is with the Department of Spine Surgery at Cath. Clinic Koblenz-Montabaur, Germany. E-mail: f.kilian@kk-km.de.
\item
 Kai Lawonn is with the University of Jena, Germany. E-mail: kai.lawonn@uni-jena.de.
}

\shortauthortitle{Eulzer \MakeLowercase{\textit{et al.}}: test}
\abstract{
We propose a visualization application, designed for the exploration of human spine simulation data.
Our goal is to support research in biomechanical spine simulation and advance efforts to implement simulation-backed analysis in surgical applications.
Biomechanical simulation is a state-of-the-art technique for analyzing load distributions of spinal structures.
Through the inclusion of patient-specific data, such simulations may facilitate personalized treatment and customized surgical interventions.
Difficulties in spine modelling and simulation can be partly attributed to poor result representation, which may also be a hindrance when introducing such techniques into a clinical environment.
Comparisons of measurements across multiple similar anatomical structures and the integration of temporal data make commonly available diagrams and charts insufficient for an intuitive and systematic display of results.
Therefore, we facilitate methods such as multiple coordinated views, abstraction and focus and context to display simulation outcomes in a dedicated tool.
By linking the result data with patient-specific anatomy, we make relevant parameters tangible for clinicians.
Furthermore, we introduce new concepts to show the directions of impact force vectors, which were not accessible before.
We integrated our toolset into a spine segmentation and simulation pipeline and evaluated our methods with both surgeons and biomechanical researchers.
When comparing our methods against standard representations that are currently in use, we found increases in accuracy and speed in data exploration tasks.
In a qualitative review, domain experts deemed the tool highly useful when dealing with simulation result data, which typically combines time-dependent patient movement and the resulting force distributions on spinal structures.
}

\teaser{
  \centering
  \includegraphics[width=\linewidth]{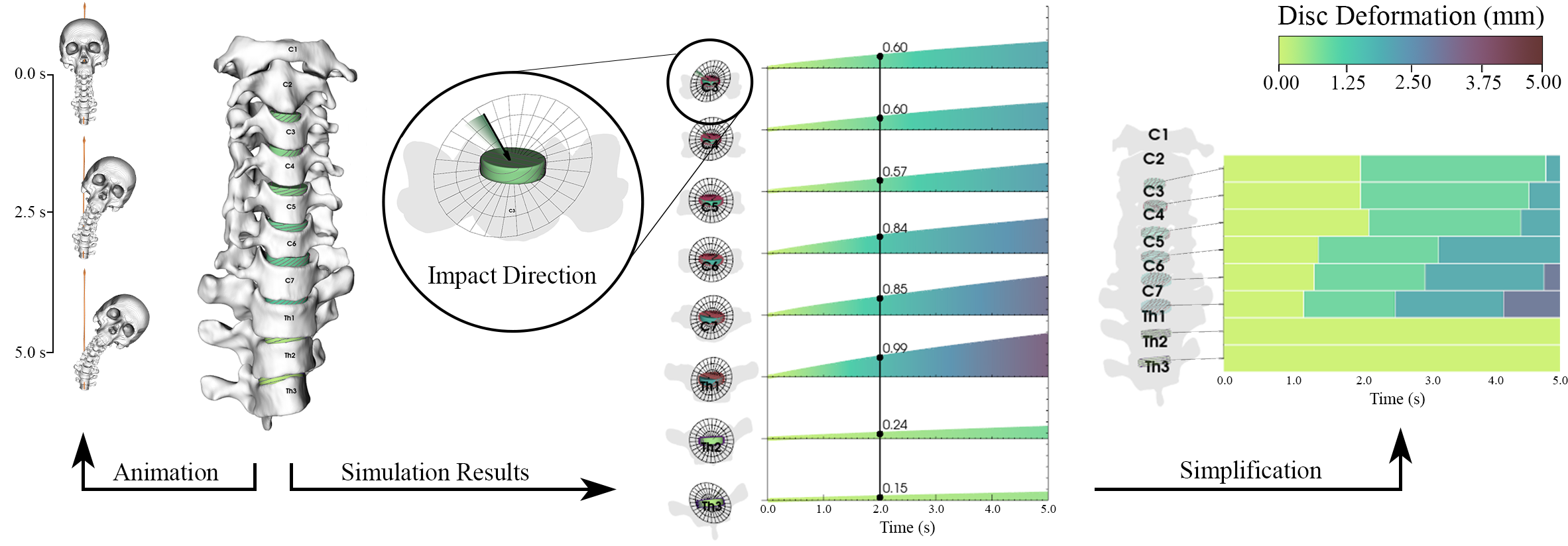}
  \caption{Overview of exploration techniques for a spinal disc deformation simulation. We extract patient movement as an animation (left), which we connect with detailed simulation results (middle), including force impact directions.
  Each value-over-time plot is associated with the corresponding spinal disc using patient-specific anatomy. A simplified depiction allows for at-a-glance assessments and ensemble comparisons (right).}
  \label{fig:teaser}
}

\begin{document}


%\firstsection{Introduction}

\maketitle
\section{Introduction}
With an estimated 80\% of the population being affected by back pain at some point in their lives~\cite{andersson1999epidemiological, who2003burden}, the prevalence of this illness has increased rapidly in recent years~\cite{gatchel2015continuing}.
Its underlying causes can range from diseases and personal anatomical characteristics~\cite{grob2007association} to modern era risk factors associated with low back or neck pain, such as overweight~\cite{mikkonen2013association, heuch2013body} or the extensive use of smartphones, resulting in a condition that received the name ``text neck"~\cite{cuellar2017text}.
Consequences are strained joints, bones and tissue, which may degenerate over time and cause chronic pain.
Treatment options are as versatile as causes, with spinal surgery being an established option to treat severe cases~\cite{rajaee2012spinal}.
%
%For instance, the annual number of performed spinal fusions, the practice of joining two or more vertebrae to restrict their movement, has increased by several magnitudes since 1998~\cite{rajaee2012spinal}.

Current research in biomechanical simulation is focused on providing a means to better understand the cause and effect relationship of spinal disorders, as well as opening up the possibility of comparing and personalizing treatment options.
Modelling forces and torques occurring in different spinal structures under various load cases can provide valuable information to clinicians and researchers.
For example, it is problematic to perform in vivo measurements to test if a procedure like lumbar fusion has negative consequences for adjacent structures~\cite{kumar2001correlation, ghiselli2004adjacent}.
A simulation, on the other hand, is certainly viable~\cite{rohlmann1999internal, bauer2015analysis}.
Similarly, more abstract research questions can be examined, such as the effects of different weight classes on spinal load~\cite{bauer2014quantification}.
Recent advances in personalized medicine, like implants build with rapid prototyping techniques~\cite{sun2005bio, rengier20103d, ay20133d}, raise the feasibility of patient-specific treatments.
Ahead of an intervention, simulation can provide valuable feedback on, e.g., force distribution, when using different implant types~\cite{bauer2017computational}.

With increasingly sophisticated biomechanical simulations emerging rapidly, the computed results are also becoming more complex.
When simulating properties of spinal structures, researchers are faced with the challenge of understanding and relating a large quantity of parameters calculated for the individual vertebrae, joints and discs.
Especially the introduction of a temporal dimension, e.g., by introducing patient movement, complicates this analysis process.
Furthermore, we will later show that medical practitioners have difficulty to understand typical simulation output, which may be a major hindrance when such systems are to be adapted in clinical environments.
We propose a visualization framework, which facilitates intuitive exploration of clinically relevant results from biomechanical spine simulation (cf. Fig.~\ref{fig:teaser}).
We integrated our methods into an existing segmentation and simulation pipeline.
Our main contributions are components of this application:
\begin{itemize}
\itemsep0.5em
    \item We facilitate exploration of all pipeline outputs: patient-specific spinal anatomy, animation of movements and temporal simulation result data.
    \item We designed anatomically aligned plots for intuitive data analysis across multiple vertebrae.
    \item We propose a 3D depiction for anatomical assessments, a 2D view for singular datasets and a simplified representation for ensemble comparisons.
    \item Additionally, we introduce embedded glyphs to encode force directions, which were not accessible before.
\end{itemize}

\section{Medical and Simulation Background}
We will now summarize necessary context and terminology from both the medical and simulation domains.

\begin{figure}[tbh]
   \centering
   \includegraphics[width=0.8\linewidth]{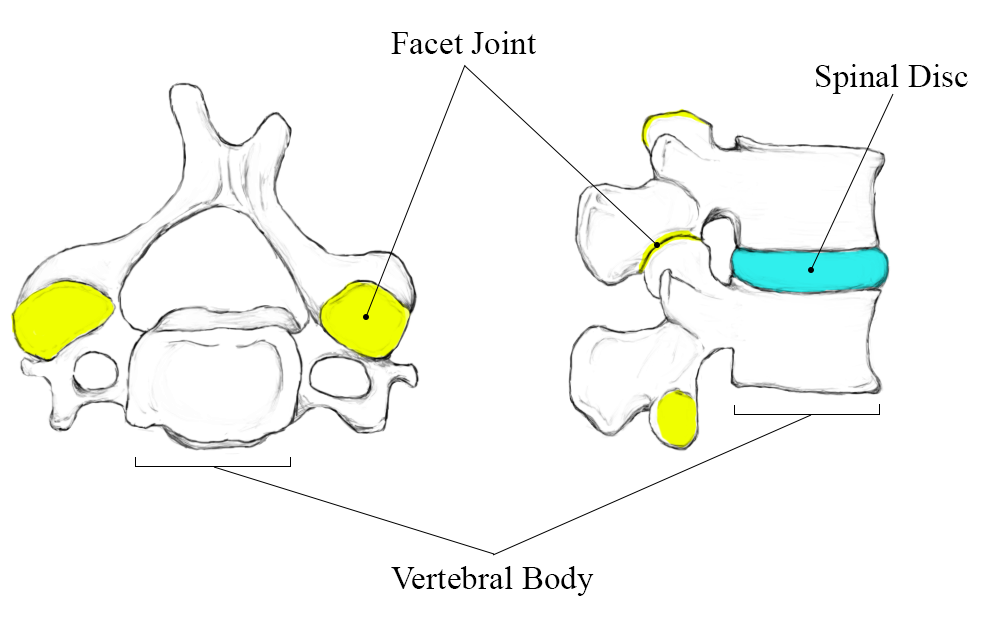}
   \caption{\label{fig:anatomy}
     A vertebra from a cranial point of view (left) and two stacked vertebrae as seen from the right lateral side (right). The vertebrae connect through two joints and a disc.}
\end{figure}

\textbf{Anatomy of the spine.} Along its length an adult's spine follows four typical curvatures, according to which the vertebrae are anatomically grouped.
From neck to pelvis, these groups are the cervical spine (vertebrae C1-C7), the thoracic spine (vertebrae Th1-Th12), the lumbar spine (vertebrae L1-L5) and the Os sacrum.
Clinically relevant are particularly the transition regions, as they are often predilection sites for spinal diseases (e.g., disc prolapse).
With exception of the first and second cervical vertebra, all vertebrae follow a common blueprint (cf. Fig.~\ref{fig:anatomy}).
The vertebral bodies are stacked enclosing discs,
%which are each made of a fibrous ring (annulus fibrosus) around a soft core (nucleus pulposus).
%
%These discs 
which cushion the surrounding bones and allow the torso to perform bending motions.
Each vertebra also connects to the structure above and below through facet joints located to the left and right of the vertebral body.
Depending on the spine segment, they are tilted in different directions, facilitating various ranges of motion.
Facets and discs are the spinal structures most prone to injury and degeneration, as they support the body's weight and distribute forces during movements.
The spine is stabilized from all sides by a complex network of ligaments and muscles, which attach to the vertebrae.

%%%%%%%%%%%%%%%%%%%%%%%%%%%%%%%%%%%%%%%%%%%%%%%%%%%%%%%%%%
%\subsection{Biomechanical Spine Simulation}
\textbf{Biomechanical Spine Simulation.} Computer simulations of spinal mechanics can provide information that helps to answer medical research questions and might become an important asset in planning surgical interventions.
Recent advances in multibody simulation allow for the modelling of patient-specific spinal structures, including the simulation of complex motion sequences~\cite{bauer2013biomechanical, bauer2015analysis, bauer2016basics}.
Typically, parameters like stiffness and damping of the modelled bodies are taken from studies performed on real tissue~\cite{panjabi1990clinical, wilke1994universal} and additional aspects can be taken into account, such as image-based tissue degeneration scores~\cite{wilke2006validity, kettler2006validity}.
A biomechanical simulation then computes the resulting forces, deformations and movements present in the scene and updates them per animation tick, i.e., minimal time-increment, starting from some initial configuration.
Researchers and clinicians are interested in the particularities of the resulting values, how they affect each structure and how they react to external factors, e.g., patient movement.
In this process, a range of difficulties needs to be addressed.
For instance, values of parameters in literature often fluctuate significantly~\cite{bauer2015does} and human spine anatomy has a high variance~\cite{roussouly2005classification}.
Therefore, comparable and intuitively understandable simulation results are of utmost importance to overcome these challenges.
\begin{figure*}[tbh]
   \centering
   \includegraphics[width=1.0\textwidth]{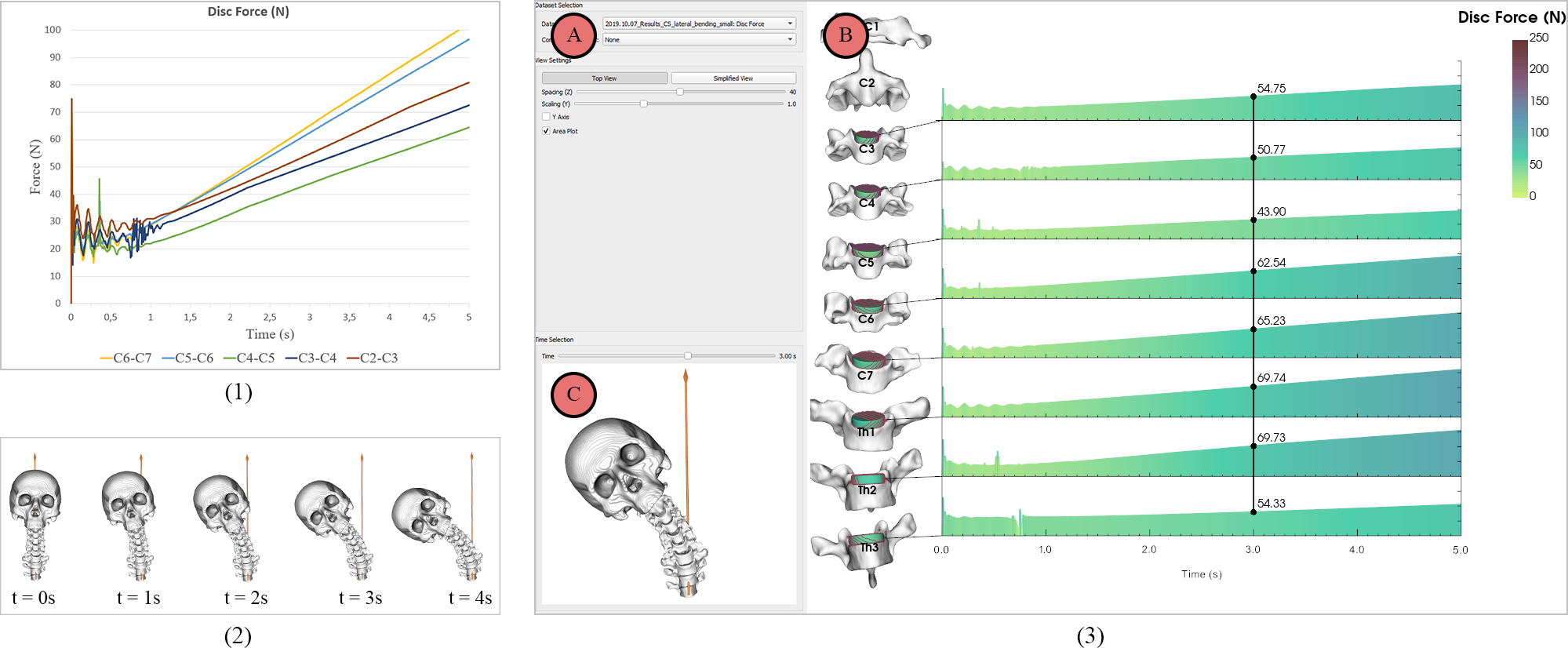}
   \caption{\label{fig:layout}
     Superposed line plots (1), a typical representation that has been used to analyze simulation results. Meaningful comparisons are hardly possible, even with only five spinal discs displayed. The same data set can be visualized with our proposed tool (3). In this example, the main window (B) shows area charts of computed parameters over time for eight spinal discs. Each chart is associated with corresponding patient anatomy. The black line highlights the selected time step and shows the plot values. Options, such as spacing between plots, can be chosen from the control panel (A). The animation window (C) links the chosen point in time with the performed movement (2).}
\end{figure*}
\section{Related Work}
Related work is comprised of the current state of the art in representation of simulation output.
Also, we summarize efforts made in visualization of anatomical features, in particular of the human spine.

%%%%%%%%%%%%%%%%%%%%%%%%%%%%%%%%%%%%%%%%%%%%%%%%%%%%%%%%%%
\subsection{Depicting Simulation Results}
A number of methods have been proposed to deal with complex data in 3D space, e.g., resulting from simulation.
% Ensemble visualization and comparative approaches
A common method is to employ ensemble visualizations and comparative approaches, which show multiple data sets within a mutual domain to indicate different values of attributes.
Potter et al.~\cite{Potter2009} proposed a framework for visual analysis of ensemble data, consisting of linked views, which present an overview first, then guide the user towards details.
%
%A drawback is that only one scalar attribute can be displayed at a time.
%
Konyha et al.~\cite{Konyha2012} explored methods for analyzing scalar attributes, which they called \textit{families of curves}.
They show different levels of data interaction through brushing, starting with single views (e.g., scatter plots), then they transition to brushing in multiple views. % allowing brushes to be combined by logical operators.
%
%While this allows an incremental exploration of the underlying data, it requires the user to keep track of a number of views simultaneously, which can be exhausting.
%
%Fr\"ohler et al.~\cite{Frohler2016} proposed a system for analysis of parameter influence on multi-channel segmentations.
%
%For this purpose, they coupled cluster visualizations and histograms.
%
Demir et al.~\cite{Demir2014} combined bar and line plots to facilitate visual exploration of 3D ensemble fields.
%
%They draw each member of the ensemble as a line chart where each 3D data point is linearized along a space-filling curve.
%
%The bar charts then depict aggregated statistical values of the line charts for defined spatial ranges.
%
They, too, display an overview first, where detail is provided on demand by narrowing the view of the camera.
Weissenb\"ock et al.~\cite{Weissenbock2018} extended this work by implementing nonlinear scaling of the 1D curves.
This leads to a more compact view, however, both depictions miss a spatial correspondence to the original volume data, where brushing is required to indicate areas of interest.

% Visualization of spatiotemporal data
A common theme in research on medical imaging and visualization are techniques for spatial comparison of 3D data, which may also be time-varying, i.e., spatio-temporal.
%
%This includes, but is not limited to, mesh comparison algorithms~\cite{schmidt2014ymca}.
%
Hermann et al.~\cite{hermann2015} demonstrated how image warping can be used to show variability in ensembles of biomedical images and Raidou et al.~\cite{raidou2018} proposed a tool for visual exploration of bladder shape variation during prostate cancer radiotherapy.
Another approach is to merge data captured at different time steps to, for example, display a static overview, allowing direct comparison~\cite{eulzer2019temporal}.
Recently, time-warping was proposed, as a method of selectively defining regions of interest around spatio-temporal events~\cite{solteszova2020memento}.
A number of methods utilized in spatial comparisons can be found in the survey of Kim et al.~\cite{kim2017}, who derive four elemental ways of visualizing such data: Juxtaposition, superimposition, interchange and explicit encoding.

% Glyph visualization
A further domain worth mentioning in this context is glyph visualization.
Different value fields in a 3D domain can be represented using glyphs placed to preserve spatial correspondence.
For example, a typical application area is visualization of tensor fields~\cite{zhang2015glyph}.
Borgo et al.~\cite{Borgo2013} defined guidelines for glyph visualizations and Ropinski et al.~\cite{Ropinski2011} introduced a taxonomy for glyphs in medical applications.
%

% Simulation visualization approaches
There exist a number of approaches to depict 3D simulation data, which generally rely on the same fundamental principles as the works above.
Krekel et al.~\cite{krekel2006interactive, Krekel2010VisualAO} showed how simulated range of motion data derived from patient-specific anatomy can be displayed using a combination of interactable 3D representations with embedded glyphs and additional statistical views.
In a pre-operative planning system, such tools can be used to assist surgeons in complex decision-making processes.
Related techniques can be found in the works of Dick et al.~\cite{Dick:2009:HipJointReplacementPlanning, Dick:2009:StressTensorFields, Dick:2011:DistanceVisualization}, who employ color maps on 3D structures, illustrative stress tensor field visualizations and glyphs indicating object distances to facilitate planning of surgical procedures.
Possible approaches are simulated, supporting clinicians in finding an optimal treatment option through visual assessment of resulting attribute values.
Depicting simulation results often becomes more intricate when the data is time-dependent.
The works of Doleisch et al.~\cite{Doleisch2003, Doleisch2004, Doleisch2007} address this problem in particular.
They utilize \textit{focus and context} techniques, where the user defines a target domain in, e.g., statistical representations, which is then visualized or highlighted in a linked 3D depiction of the simulation.
%
%They show application areas in computational fluid dynamics, where the temporal aspect of the underlying data is usually depicted through animation or iteration.
%
Similar cases have been explored in the medical domain when simulating blood flow ~\cite{Meuschke:2017:GCS:3128397.3128407, Meuschke_2017_TVCG, Lawonn_2015_TVCG, Lawonn_2014_CGFb}.
Glyphs, color maps, illustrative techniques and integration of 3D models and statistical data representation methods, such as plots and charts, are used to visualize a range of parameters within the 3D domain.
This means the anatomical structure can be used to enable a direct link between attribute values and a patient's anatomy.
For an overview of illustrative visualization techniques and focus and context depictions, we refer to the surveys of Lawonn et al.~\cite{doi:10.1111/cgf.13306,Lawonn_2018_CGF,Lawonn_2015_Feature}.

%%%%%%%%%%%%%%%%%%%%%%%%%%%%%%%%%%%%%%%%%%%%%%%%%%%%%%%%%%
\subsection{Visualizing Properties of Spinal Structures}
Patient-specific anatomical features are commonly explored using cross-sectional imaging modalities, which in turn can serve as the basis for volume renderings or segmented 3D models.
%
%Recently, several methods emerged, which utilize dimensional reduction techniques, i.e., flattening or reformation, thus reducing the number of dimensions that need to be navigated during exploration~\cite{kreiser2018}.
%
%An example would be the technique shown by Kretschmer et al.~\cite{kretschmer2014}, who generate flat representations of anatomical features, facilitating a comprehensive overview that can be used to support diagnosis, navigation or annotation.
%
%In spite of these trends to simplify and enhance analysis of patient-specific anatomical features, the particular domain of spinal structure visualization has only been sparsely explored.
%
The particular domain of spinal structure visualization, however, has only been sparsely explored.
Notable are the works of Klemm et al.~\cite{klemm_2013_VMV, klemm_2014_tvcg, klemm_2015_ivapp}, which evolve around visual analysis of lumbar spine cohort data sets.
They use a semi-automatic detection of the lumbar spine from volume images, resulting in 3D models that serve as a foundation for advanced processing.
For instance, they can be used to visualize spinal canal variability in cohort study data, allowing to draw associations between anatomy and demographic or biological factors.
Klemm et al. demonstrate how this can be achieved through clustered 3D streamlines and also geometric abstractions that only require a 2D representation~\cite{klemm_2013_VMV}.
Later, they generalized their methods towards a visual analytics workflow, allowing epidemiologists to generate and validate hypotheses.
The lumbar spine cohort visualization methods served as a demonstration~\cite{klemm_2014_tvcg}.
Further, they showed how patient-specific properties can be measured and visualized using geometric spine models.
These can be used to analyze mutual dependencies between shape-describing parameters and other variables, allowing new insights into spine data sets~\cite{klemm_2015_ivapp}.

To the best of our knowledge, there have been no works specifically targeting visualization of spine simulations.
Common mechanical simulation tools~\cite{fishwick1992simpack, delp2007opensim} generally offer a result representation based on standard plotting, e.g., line or bar charts, which can be inadequate for understanding complex systems like spinal anatomy.

\begin{figure*}[tbh]
   \centering
   \includegraphics[width=1.0\textwidth]{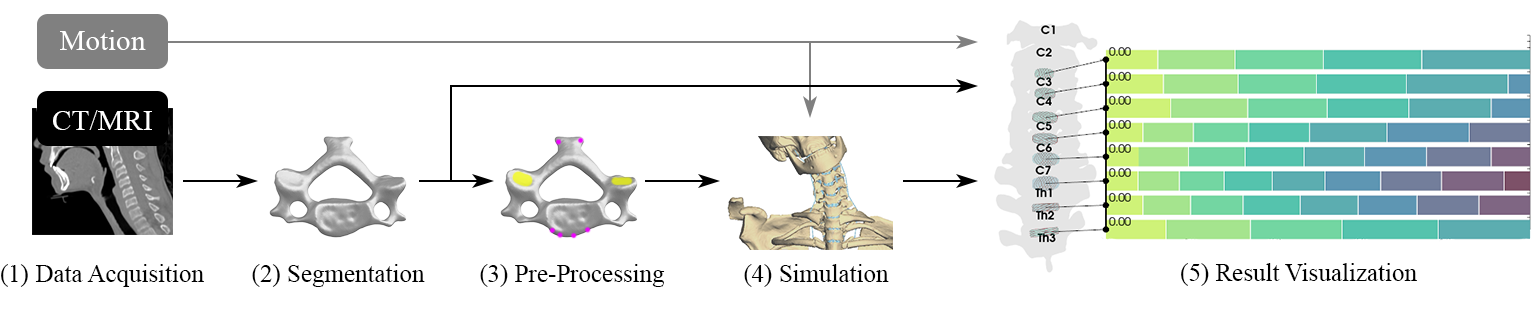}
   \caption{\label{fig:pipeline}
     The full pipeline. From medical volume images (1) vertebra models are segmented (2). Then, they undergo pre-processing where anatomic markers, e.g., origin and attachment of ligaments are detected (3). Based on this information the biomechanical simulation is performed (4). In this work, we show how the results can be visualized (5). We target to facilitate inclusion of patient-specific motion data, which is currently being integrated into the pipeline.}
\end{figure*}
\section{Requirement Analysis}
We worked closely with domain experts when distilling requirements for our framework.
Our intention is to provide a useful tool for both spinal simulation researchers and medical practitioners, as we believe making the output data intuitively understandable could benefit either and may prove useful to bridge the gap between technical research and clinical application.
For this reason, we consulted an expert in each field.
In the domain of biomechanical simulation we worked with a researcher who has 15 years of experience in spine simulation.
We then discussed possible directions with an orthopedic and neurosurgeon, specialized in spinal surgery with 36 years of experience.

%%%%%%%%%%%%%%%%%%%%%%%%%%%%%%%%%%%%%%%%%%%%%%%%%%%% new
% what are current workflows like?
As a first step, we assessed the current workflows applied in biomechanical spine simulation research.
In patient-specific simulations, segmentations of vertebra geometry are created based on common medical volume imaging, e.g., computed tomography (CT).
Pre-processing prepares the models for spine simulation, for instance, ligament attachments and origins can be detected and marked.
Then, the model's biomechanical properties are simulated, which includes forces, displacements, and deformations.
%
% whats the data that is output?
This results in attributes computed per simulation tick, i.e., smallest time increment, for each anatomical structure over a specified period of time.
%
% how is it interpreted?
Until now, these results were displayed using line charts, for instance, as total force over time plots for selected structures.
In current workflows, multiple structures are compared by using superpositions of several line charts (Fig.~\ref{fig:layout} (1)).
% why is this not optimal?
This only allows for a limited number of comparisons and quickly results in a cluttered view.
Another problem we found in discussion with physicians is that they have difficulties to grasp anatomical correspondence when presented with these kinds of representations.
As soon as multiple line charts were displayed, they were hesitant when talking about the underlying anatomical structures and sometimes even referred to wrong ones.
This not only made the clinical experts reluctant to implement such a simulation system into clinical routine but could also be potentially harmful.
%

% which data is relevant?
To provide an improved representation of the simulation results, tailored to the needs of researchers and clinicians working in this domain, we first narrowed down which simulation attribute values are of importance.
Medically relevant are particularly force distributions on spinal discs and facet joins, as they are often the sources of chronic pain.
This can be a result of unusually high or unbalanced forces, which are mechanical conditions that can be appropriately simulated.
Other important clinical parameters are the resulting deformations of the spinal discs.
These are typically measured using imaging techniques and can be simulated through model stiffness, based on factors like degree of degeneration and patient age.
Visualizing these aspects might contribute to a better clinical analysis, for instance, when pathological cases can be more accurately identified and classified and adequate treatment options can be reviewed or even simulated.

% what are tasks that experts want to perform?
In our discussions, we attempted to define the data analysis tasks researchers and clinicians wish to perform and from these tasks specified a number of requirements a visualization framework would need to fulfill.
Both biomechanical researchers and physicians, desire an overview of individual data sets, i.e., distribution of a selected attribute over multiple vertebrae.
In addition to the simulation output, the experts agreed that a display of the segmented geometry is of major importance, as data interpretation is highly dependent on the specific anatomical features.
In order to identify imbalances, they want to directly compare forces acting on the left and right facet joints.
Furthermore, it would make sense to allow for comparison of multiple data sets.
While physicians foremost intend to compare different simulated treatment options, biomechanical researchers would also like to contrast distinct sets of input parameters.
The display of total forces is particularly significant, however, the experts found it promising to also incorporate force \textit{direction} on spinal discs, as vertical forces can be more easily compensated, while non-orthogonal or shear forces can lead to injury.
Moreover, the experts agreed that an exploration of the temporal dimension of the simulation data is critical.
A current research goal of biomechanical spine simulation is to account for movement patterns of the patient.
This is also an important clinical aspect, since spinal structures are essential for almost all daily motions and load distribution is dependent on the way movements are performed.
%
%While statistical range-of-motion data has been considered when evaluating models~\cite{cook2015range}, patient-specific motion data is hard to incorporate.
%
%$This is in part due to the difficulty of result analysis.
%
%For instance, forces on a number of spinal discs in a static model can be compared with juxtaposed bar charts~\cite{bauer2014quantification}, however, a temporal dimension cannot be trivially added.

% what are tasks that only one expert group wants to perform?
While many potential exploration tasks seem similar between our target groups, we also encountered some differences.
When examining values, physicians tend to be less interested in exact numerical output and more in averages and spikes in data.
Especially, they require irregular force patterns to be easily discernible.
For researchers, on the other hand, reading out exact values is a necessity, since the results of different models need to be compared quantitatively.
They also need to quickly identify faulty or missing data.
% Do these differences in tasks result in complications?
In a visualization framework, we believe these differences could be accounted for using optional features that can be chosen according to the task.

% what specific requirements did the experts mention?
The experts expressed specific requirements, when we discussed how a visualization framework should facilitate the gathered data exploration tasks.
Especially for medical practitioners the relation between the data and anatomical structures should be intuitive, e.g., the connection between a force over time plot and the corresponding spinal disc should be clear.
Both experts argued that intuitive comparisons are most crucial for similar structures, for example neighboring spinal discs or left and right facet joints.
To facilitate understanding of the temporal dimension, we found a direct connection with the movement that is actually performed to be a requirement.
Last but not least, we concluded that the important component of a force direction visualization would be to encode how vertically forces are impacting each spinal disc or whether shear forces are present.
As many tasks and requirements are aligned, we believe a framework targeting both user groups is a promising direction.
Ultimately, it could also form a common basis for communicating results.
This is why we propose a visualization framework to facilitate the following summarized requirements:
\begin{description}
\itemsep0.5em 
    \item[R1] There should be a clear correspondence between simulation result data and patient anatomy.
    \item[R2] Intuitive comparisons of result values should be possible across similar structures.
    \item[R3] The displayed data is intrinsically time-dependent, requiring an explicit connection between patient movement and attribute values.
    \item[R4] The directions of forces impacting a patient's spinal discs should be accessible.
\end{description}

\section{Methods}
We integrated our visualization tools into a state-of-the-art pipeline for patient-specific biomechanical spine simulation (Fig.~\ref{fig:pipeline}).

The pipeline involves an automatic segmentation process~\cite{al2018automatic}, followed by data pre-processing, and a multi-body-simulation~\cite{bauer2013biomechanical, bauer2017computational}.
%
%The visualization, however, is not dependent on this particular simulation method and may well be used in other scenarios.
%
Biomechanical researchers are currently incorporating patient-specific spinal motion data, captured with visible light techniques, into the simulation model.
This would allow a clinical analysis of both individual anatomy and movements, in a combined system.
For the visualization of the results, we attempt to integrate all components of this pipeline, which are relevant to data analysis.
We, therefore, combine the numerical results, spine model, and motion data into one framework.
As we cannot utilize real motion data yet, we use artificial head movements by adding external forces to the simulation.
This demonstrates the functionality and allows us to evaluate the tool, making it useful to simulation experts already.

While our proposed system can be implemented with an arbitrary amount of vertebrae, we decided to use a model of the cervical spine first.
In clinical procedures and analysis, as well as property modelling, often the cervical, thoracic, or lumbar spine (and the corresponding transition areas) are focused.
This allows for a more targeted point of view.
In the following, the used model consists of vertebrae C1-C7 and the transition to Th1-Th3.
Our methods should be similarly applicable to the thoracic or lumbar spine.

%%%%%%%%%%%%%%%%%%%%%%%%%%%%%%%%%%%%%%%%%%%%%%%%%%%%%%%%%%%%%%%%%%%%%%%%%%%%%%%%%%%%%%%%%%%%%%%%%%%
%
\begin{figure}[tbh]
   \centering
   \includegraphics[width=1.0\linewidth]{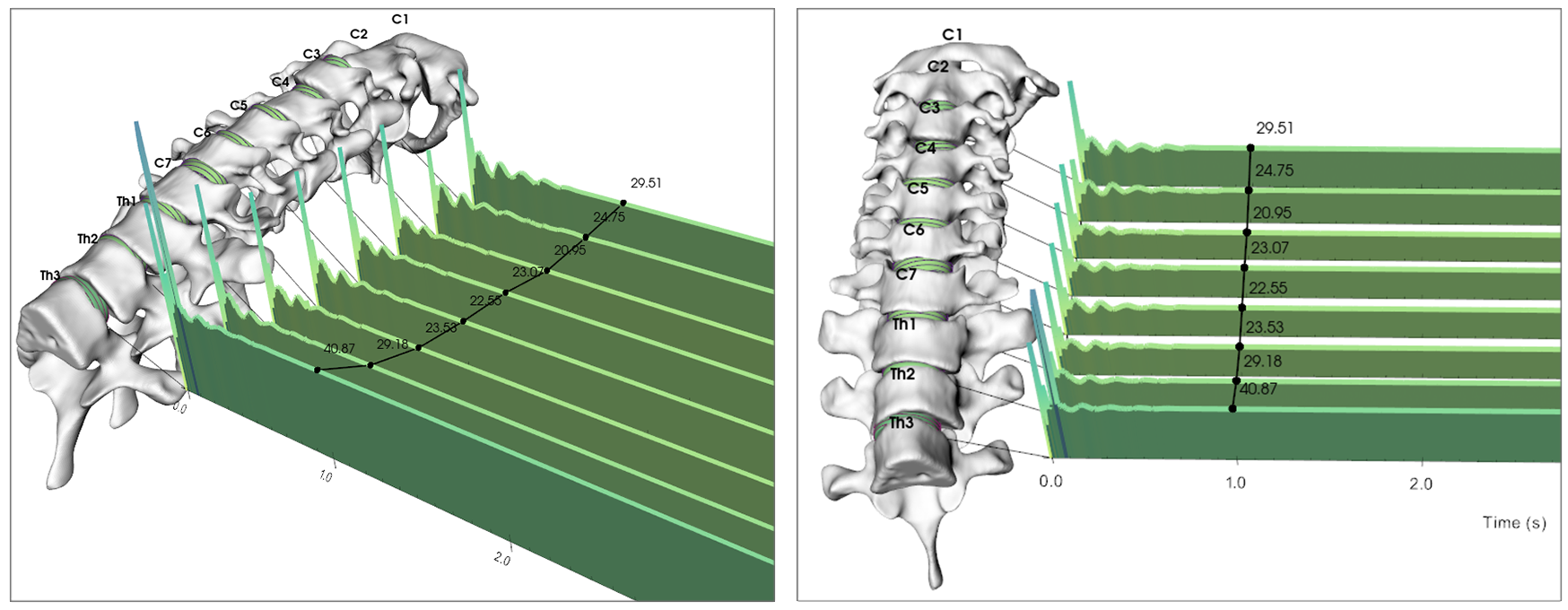}
   \caption{\label{fig:3D}
     A 3D depiction allows free interaction with the patient's spinal anatomy, while the stacked charts remain linked to their target structure.}
\end{figure}
\begin{figure}[tbh]
   \centering
   \includegraphics[width=1.0\linewidth]{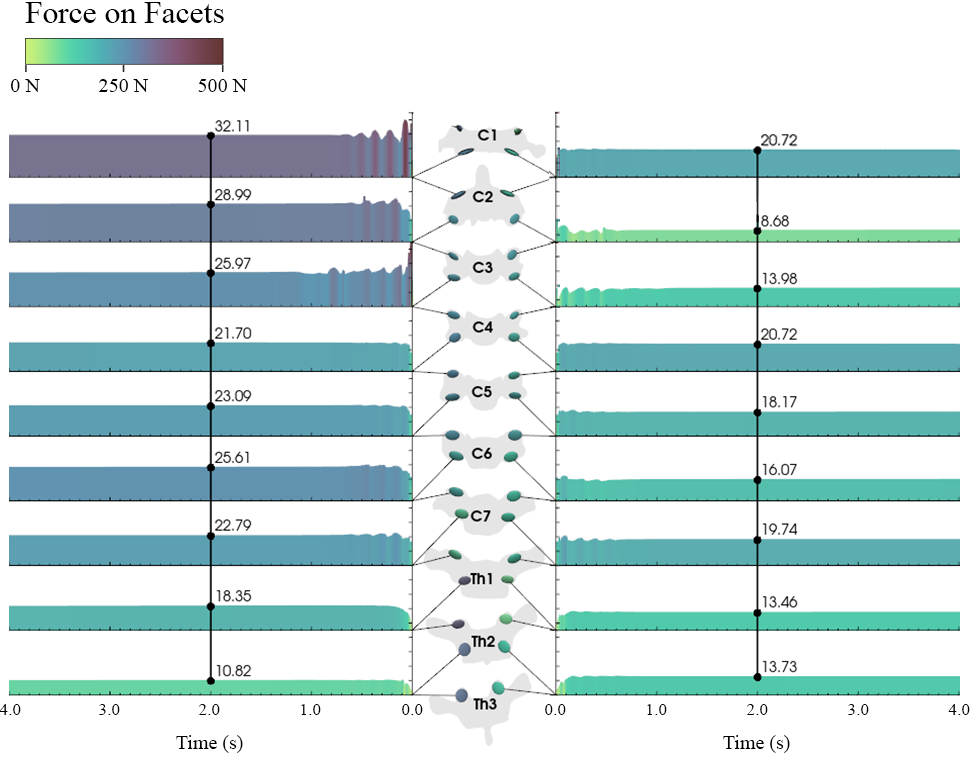}
   \caption{\label{fig:facets}
     Values computed on left and right facet joints are shown on each side respectively. The left time axis is flipped to achieve a mirror-symmetry effect and improve comparability of the two sides. Note how this patient appears to have an imbalance in force distribution across the first facets (C1-C4).}
\end{figure}
\subsection{Choice of Data Representation}
We output raw data from the simulation in form of large matrices, one per observed value, e.g., force on spinal discs.
Each row corresponds with a simulation tick interval, while every column represents a structure, for instance a particular disc.
This means, we can extract a value like force or deformation as a function of time per focus structure.

\begin{figure}[tbh]
   \centering
   \includegraphics[width=1.0\linewidth]{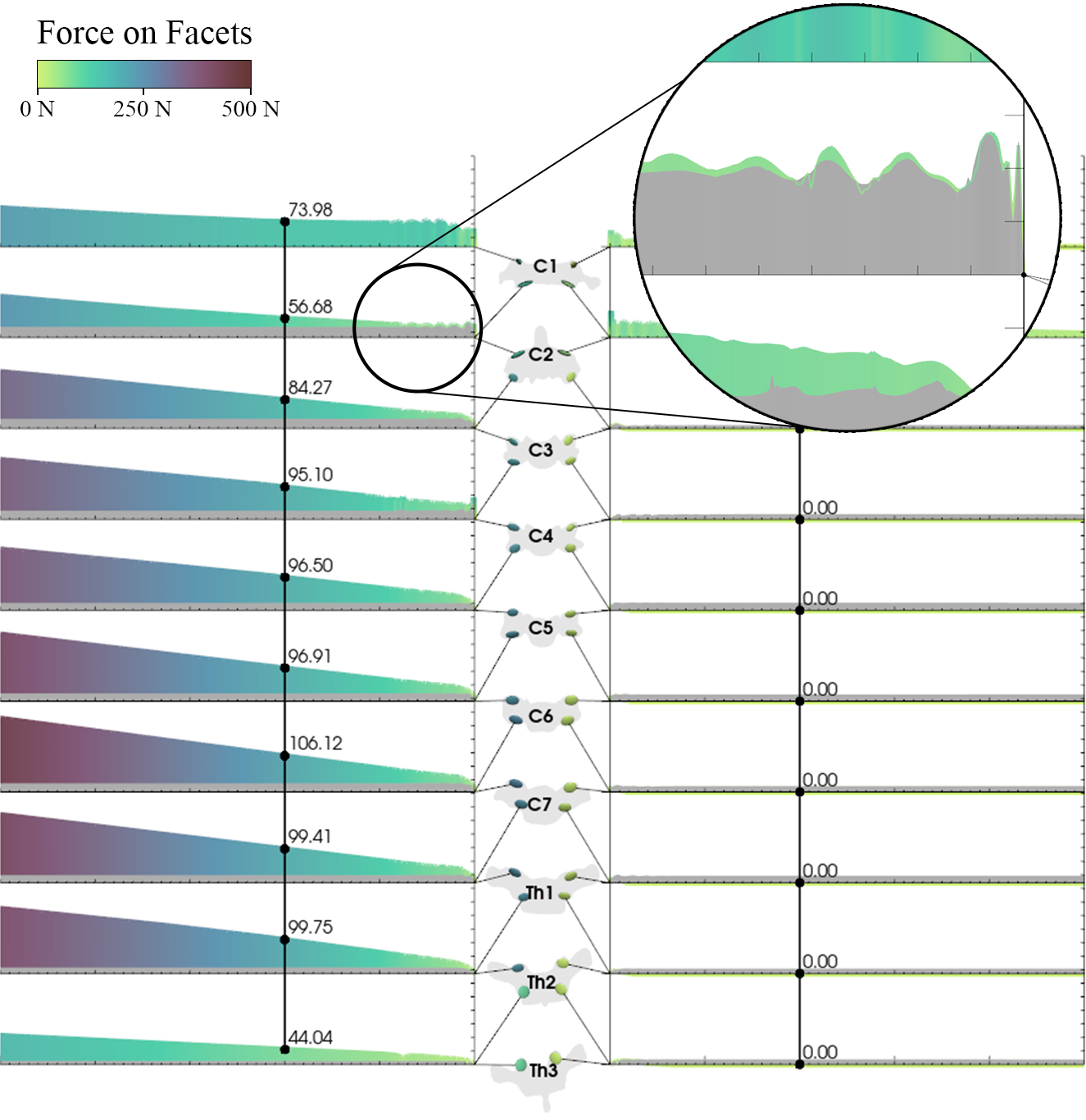}
   \caption{\label{fig:comparison}
     This data set shows facet force distributions following a lateral head bend. It is directly compared against a static simulation without movement (gray area plot). It can be seen how tilting the head results in a decisively higher load on one side of the vertebrae.}
\end{figure}
Fundamentally, multiple value-over-time plots can be displayed side-by-side (juxtaposition), within a common coordinate system (superposition), or through interchange of a selected plot.
The latter is not desirable, as we intend to address the comparability requirement (\textbf{R2}).
This leaves two main options: using a single coordinate system or one for each plot.
A superposition in two dimensions, as it is used in current workflows, is not suited to compare a high number of plots.
A possibility would be to stack the charts in a third dimension.
This is generally not an ideal method, as it requires direct scene interaction from the user and assessments may be inaccurate due to perspective distortion.
Still, we implemented this layout as an option, since it showed to have an advantage when addressing \textbf{R1}: the anatomical context of the data can be displayed in the 3D domain, directly next to the corresponding charts (Fig.~\ref{fig:3D}).
The spine can be rendered in its anatomically correct state and freely rotated, while correspondence to the simulation result data remain clear.
The domain experts found this representation helpful to gain a first impression of the combined data and anatomy.
To address the shortcomings of this depiction, e.g., possible perspective distortion, we implemented the second option as the default view to explore the result values.
We use juxtaposed charts, aligned on a shared time-axis, which we display in 2D (Fig.~\ref{fig:layout} (3)).
Comparing area and line charts, it quickly became apparent that area charts are better suited, as they intuitively convey value dimensions, even with many charts in one scene.
We can still keep the anatomical context, by aligning the charts beside the vertically drawn spine geometry and keeping datapoints right next to their respective anatomical structure.
The drawback of this representation is that the geometry cannot be interacted with freely, without losing correspondence.
Another challenge is that the data may encompass a wide range of values, with some plots filling their coordinate systems while others remain close to zero.
To enable comparison, all axes need to be equally scaled, but a low overall scale may impact assessment of small plots.
We solve this problem by giving the user the option to adjust the distance between plots, i.e., the length of the plots' value-axes.
In order not to lose anatomical context, we expand the spine geometry accordingly by pulling the vertebrae apart, i.e., translating them along the vertical axis.

We determined that users need to be able to read out quantitative values and also intuitively understand general value ranges.
There are several ways to achieve this, with the simplest being to use labelled value axes for each plot with, e.g., gridlines.
We tested this option but decided to leave it disabled per default.
With around ten juxtaposed charts the scene becomes cluttered and values are hard to read.
Instead, we opted for a combination of colormapping and point selection.
We employ a consistent colormap of data values across all views, which targets to visualize the approximate value ranges and help with general assessments.
We use the viridis colormap, as its colors are perceptually uniform and can also be perceived by most forms of color blindness~\cite{liu2018somewhere, ware2018measuring}.
To see quantitative values for each chart in the scene, the user may select a point in time of the simulation via a continuous slider (cf. Fig.~\ref{fig:layout} (B)).
\begin{figure*}[tbh]
   \centering
   \includegraphics[width=1.0\textwidth]{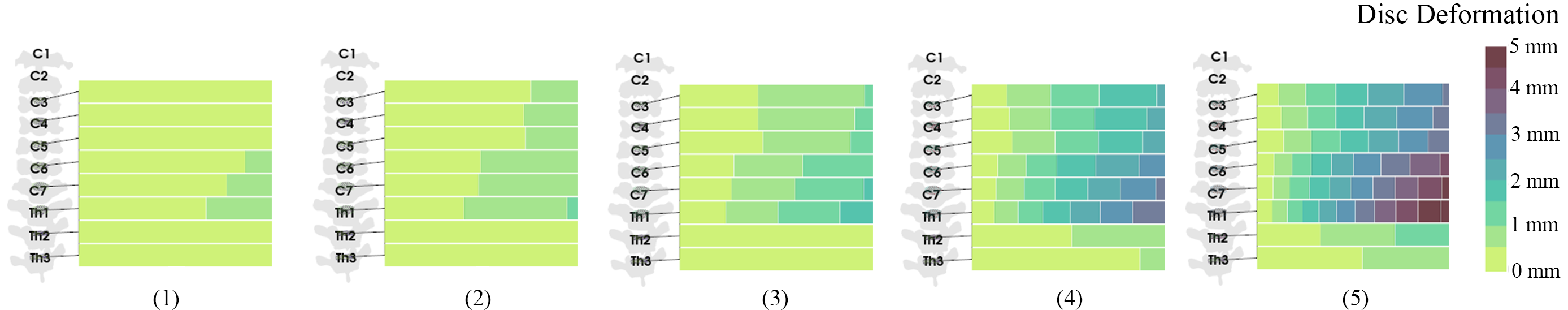}
   \caption{\label{fig:simplified}
     The simplified representation relies on a color-only encoding and facilitates comparison of simulation results over multiple data sets. Here, we show how this could be utilized to compare different input parameters. The same motion was used to simulate spinal disc deformations with an increasing degeneration degree from (1) to (5). The discretized colormap was added in a later iteration and allows to quickly discern value ranges. For instance, discs with degrees (1) to (3) appear to deform less than 2 mm.}
\end{figure*}
%%%%%%%%%%%%%%%%%%%%%%%%%%%%%%%%%%%%%%%%%%%%%%%%%%%%%%%%%%%%%%%%%%%%%%%%%%%%%
\subsection{Displaying Facet Joint Data}
Spine simulations contain various structures of medical interest.
Spinal discs are a typical example, but as described above, facet joints can be similarly important when evaluating stress distributions.
In our proposed layout, spinal disc data can be shown in the standard way of right-facing plots.
Facet joints, however, pose an additional challenge, as two facets exist between each vertebra pair.
For values corresponding to facets, we therefore display right- \textit{and} left-facing plots, with the patient's spine acting as a central vertical axis (Fig.~\ref{fig:facets}).
Our main goal regarding facet data is to facilitate left-right comparisons, i.e., to visualize whether a load is equally distributed on both sides.
To enhance perceptual recognition of differences between the two sides, we apply the Gestalt Principle of Symmetry~\cite{freyd1984force}.
Use of symmetry has been shown to improve readability and understanding, for instance in graph layouts~\cite{bennett2007aesthetics, van2008perceptual}.
Thus, we create a reflective symmetry between values on the left and right side, by mirroring the time-axis of the left-hand data charts.
This results in a view, in which for all charts early points in time are closer to the central axis (where the vertebrae are rendered) and later points in time are further away.
%
%This allows a direct comparison of the two sides, addressing the equality of load distribution on the left and right side of the spine.
%
%Regardless of the focus structure, the user can enable a second data set to be shown as gray plots.
%
%This can be used, e.g., to compare a dynamic against a static simulation or contrast results of different surgical options (cf. Figure~\ref{fig:comparison}).
%
%We draw the gray comparison plot in front of the existing area.
%
%In case of occlusion, i.e., when the comparison plot value is higher, the original data set is shown as a line within the gray area.
%
%%%%%%%%%%%%%%%%%%%%%%%%%%%%%%%%%%%%%%%%%%%%%%%%%%%%%%%%%%%%%%%%%%%%%%%%%%%%%
\subsection{Movement Integration}
Up until now, the time-dependency of the load distribution can be observed in direction of the time-axis.
To connect these results with the underlying movement (\textbf{R3}), we show an animation of the model's rotations and translations, which can be derived from the simulation in the same format as the value matrices.
As the already displayed vertebrae models serve as anatomical references, using them in an animation would result in a tangled view without clear correspondences.
Therefore, we show the animation in an additional smaller window (as in Fig.~\ref{fig:layout} (C)), which is also interactable, in case different viewing angles are desired.
If animation data was generated, the time point selection is automatically linked to the respective animation time step.
This is indicated in the main window by the black selection line, which also shows the individual plot values (cf. Fig.~\ref{fig:layout} (B)).

%%%%%%%%%%%%%%%%%%%%%%%%%%%%%%%%%%%%%%%%%%%%%%%%%%%%%%%%%%%%%%%%%%%%%%%%%%%%%
\subsection{Multi-Set and Ensemble Visualization}
In some application scenarios it might be desirous to perform comparisons across multiple data sets or ensembles.
For instance, in simulation research, this applies to contrasting different initial configurations or parameter sets.
In medical practice, the results of possible treatment options or implant types on force distributions are ideally directly comparable.
Data set ensembles may also arise from cohort studies evaluating if anatomical factors contribute to the manifestation of certain pathologies.
Even simple tasks, like the comparison of two movement patterns, require more than one data set to be shown in the scene.
Especially the latter case can be covered by rendering a second simulation outcome within the existing coordinate systems.
In our tool, such a data set can be displayed in gray on top of the original area charts (Fig.~\ref{fig:comparison}).
In case of occlusion, i.e., when the comparison plot value is higher, the original chart is shown as a line within the gray area.

For ensembles we need to apply a more scalable solution, which we will call a simplified view.
To facilitate coherence, our approach retains the general layout, i.e., the spine anatomy stays as a contextual reference in the center and the value-over-time graphs are presented towards the left (and right if necessary).
As the number of charts goes up, the positional encoding of area plots is increasingly hard to read.
Therefore, we reduce the data value encoding to color only.
This allows for a quick overview and comparison of many data sets (Fig.~\ref{fig:simplified}).

%%%%%%%%%%%%%%%%%%%%%%%%%%%%%%%%%%%%%%%%%%%%%%%%%%%%%%%%%%%%%%%%%%%%%%%%%%%%%
\subsection{Force Direction Visualization} \label{Force Direction Visualization}
Standard charts allow an interpretation of a total value or its components over time.
For instance, a force vector's length or one of its $x,y,z$-components can be interpreted.
This makes it effectively impossible to understand from which spatial direction a force is impacting an anatomical structure.
Since we render a 3D representation of the patient's spine already, we propose the integration of markers showing the direction of forces within the scene to address \textbf{R4}.

\begin{figure*}[tbh]
   \centering
   \includegraphics[width=1.0\textwidth]{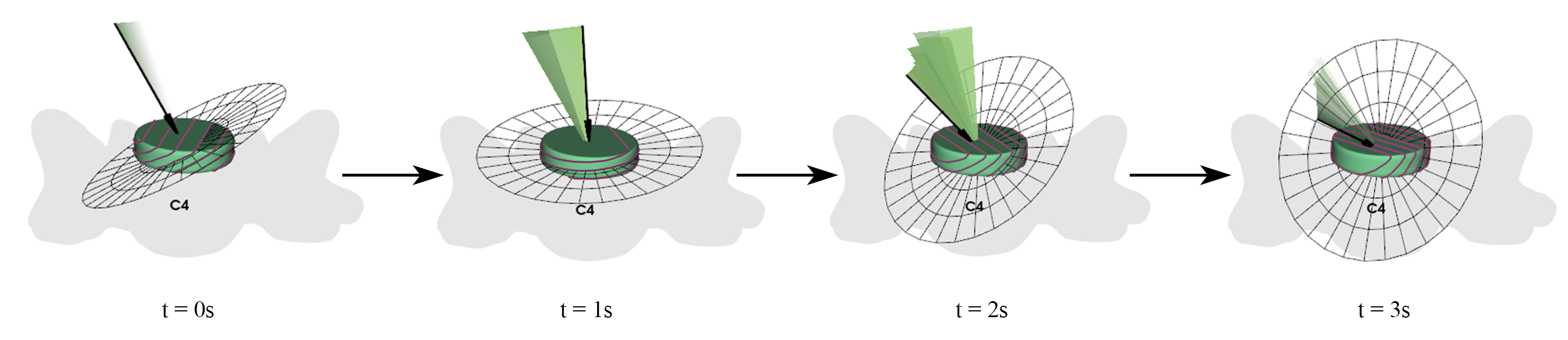}
   \caption{\label{fig:glyphs_shear}
    The directional glyphs are animated according to the time-selection. In this example, the impact originates from the top-left side (0s), then moves to a vertical direction (1s) and lastly results in a left-frontal shear force (3s). After the evaluation we improved the representation by drawing the trajectory of the arrow glyph as a fading surface. This makes the jittering of the impact direction at around 2s visible, even without an active animation.}
\end{figure*}

We display arrow glyphs as a simple and intuitive shape to represent force vectors.
Each glyph targets the barycenter of its focus structure, e.g., spinal disc model, akin to the internal representation of forces in the multi-body-simulation.
This is also in accordance with the feature-driven glyph placement approach typical for medical applications~\cite{Borgo2013}.
The arrow direction is extracted according to the selected point in time.
An option might be to scale the arrow glyphs proportional to their total force, but we propose to keep their length uniform, as this leaves less ambiguity w.r.t. their spatial direction.
Even then, the direction of arrow glyphs is difficult to interpret in 3D and depends on the viewing angle.
%
%Thus, we attempted to improve the representation to more specifically target its use-case.
%
Consulting with our domain experts, we found the most important aspect to be a clear indication of whether shear forces are present, i.e., if force vectors are impacting spinal discs more from the sides than above.
%
%Additionally, regarding the temporal dimension, we were encouraged to visualize if impact directions behave relatively stable or whether they jitter.
%
%Both aspects are currently inaccessible and cannot contribute to data assessment.
%
\begin{figure}[tbp]
   \centering
   \includegraphics[width=1.0\linewidth]{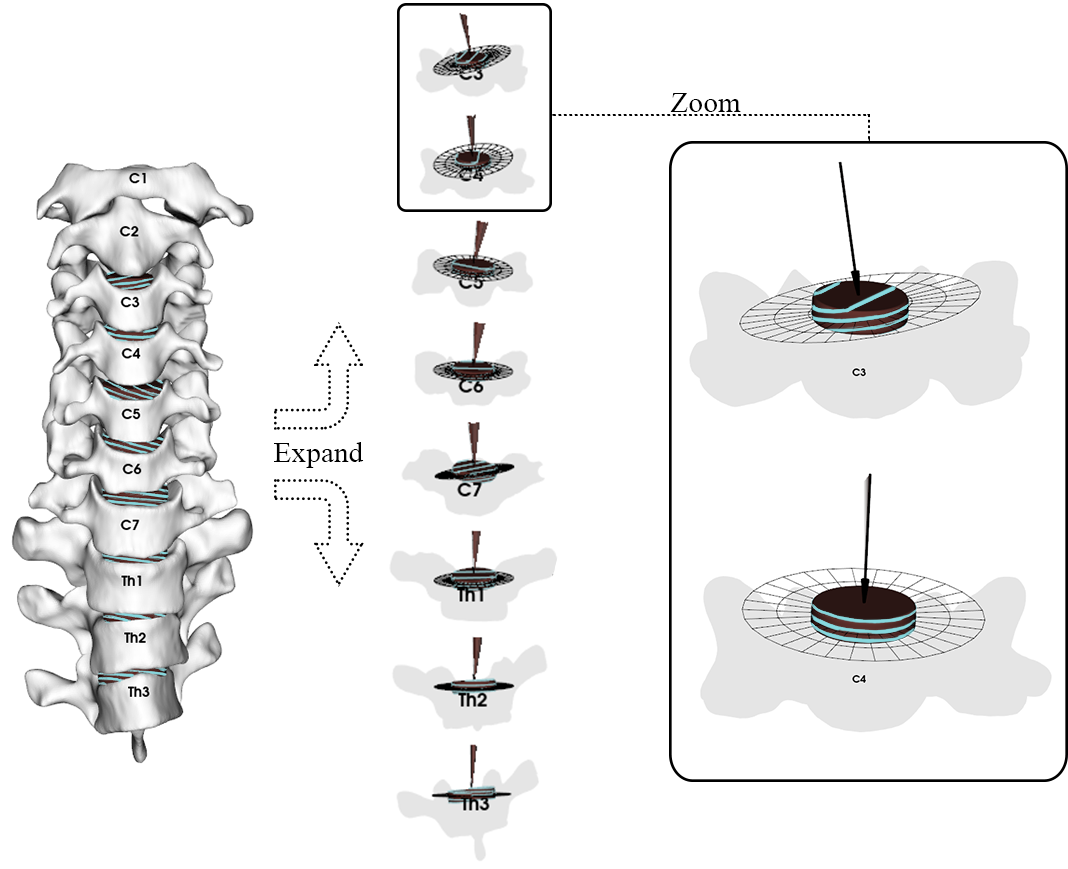}
   \caption{\label{fig:glyphs}
    Glyphs encoding the direction of forces impacting the spinal discs. They become visible through expansion of the spine geometry, which is simplified to a silhouette representation. The selected point in time is marked with an arrow and a black disc in the ``force plane'', i.e., orthogonal to the impact direction. After the evaluation, we also added isolines on the spinal disc surface, which are rendered parallel to this plane, allowing a better interpretation of how vertically the load is distributed on the spinal discs.}
\end{figure}
Therefore, to make the arrow orientations better comparable to the respective spinal disc, we propose to add an orthogonal disc located at the arrow tip (Fig~\ref{fig:glyphs}).
It can be thought of as a ``force plane'', with the impact vector being its normal.
This planar component simplifies interpretation of the direction and gives spatial cues, even when little to no 3D interaction is used.
The orientation of the glyph disc can be visually compared against the respective spinal disc, giving an impression of how vertically the load is distributed at the selected time interval.
To avoid visual clutter, these glyphs are only shown when the vertebrae models are pulled apart (through the axis scaling adjustment) and disappear when the vertebrae are condensed towards their anatomically correct ``stacked'' positions.
All glyph components are updated according to the time-point selection, resulting in an interactive animation of force directions (Fig.~\ref{fig:glyphs_shear}).

One obstacle to keep in mind is that occurring forces in a simulation are possibly not measured in the local coordinate system of the target structure.
In our case, each force vector $\mathbf{f}$ is extracted in global coordinates, while the target object's position is determined by some rotation matrix $\mathbf{\phi}$, followed by a translation.
As we display the direction glyphs in a model with fixed orientation, we need to correct the force vector's orientation by $\hat{\mathbf{f}} = \mathbf{\phi}^T \mathbf{f}$, before positioning the glyph in the scene.
\section{Evaluation}
During development of our methods we iterated several feedback cycles with domain experts.
We will now describe the main evaluation of our initial prototype. % which did not yet include the final functionality as presented so far.
This evaluation was particularly beneficial, as it yielded some relevant design decisions.

We conducted individual interviews with six domain experts (1 female, 5 male; 26-63 years old; median 41 years). 
The expert group can be divided into three biomedical simulation researchers (5, 2 and 3 years of experience) and three physicians specialized in spine surgery (16, 20 and 36 years of experience as orthopedics or neurosurgeons).
None of the experts had used the proposed tool before.
Also, they were not part of the preceding discussions, except for one of the interviewed surgeons, who had also been present for the original requirement analysis.
Participants were introduced to the concept of the full acquisition, simulation and visualization pipeline.
We showed them examples of superposed result plots and explained how these are currently used to interpret the simulation output.
Then, they were presented with the same data sets in our framework and we introduced them to all features, while they were able to freely explore the data.
We encouraged participants to think aloud while they interacted with the tool and noted down their spoken comments and suggestions, particularly any difficulties they encountered.
In an attempt to recreate scenarios that might arise during typical use of the tool, we integrated a number of tasks in this process.
%
%To probe if our visualizations increased the experts' exploration capabilities, we alternated between old representations and our tool, i.e., as in Fig.~\ref{fig:layout} (1) and (3).
%
Participants wrote their answers on paper, while interacting with all data representations on a single-screen laptop setup.

The first task targeted simple readouts of values at predetermined points in time of the simulation.
This is a common task for simulation researchers, who need to perform quantitative measurements and comparisons.
For each value, participants were to indicate how confident they felt with their statement on a scale from $0\%$ (very uncertain) to $100\%$ (highly certain).
To compare our tools with former representations, participants conducted the task twice, once using the old representation and once with our visualization framework, i.e., as in Fig.~\ref{fig:layout} (1) and (3).
For comparison, we used the same data set, however, we specified different time points to avoid a learning effect.
We also switched the order of methods for each new participant, i.e., whether the old representation or our framework was used first.

The second task was conducted similarly, but focused on general impressions and assessments that participants were able to draw from the shown data.
We matched our questions to theoretical exploration goals of clinicians, which we gathered during the requirement analysis.
For instance, we asked participants whether they found the displayed disc deformation value for a number of determined spinal discs to be higher or lower on average, as compared to the rest.
We also used facet data sets, where for each facet pair between two vertebrae participants were to decide if load distribution was skewed to the patient's left or right side or whether it was approximately equally balanced.
They could also note down missing data sets (we removed data from one facet joint) and mark their certainty, as before.
Again, the participants performed the task twice, once with the new and once with the old data representation.
%
%In the latter scenario we gave them two charts to compare the left and right side of facets.

To determine the effectiveness of the directional glyphs, we asked the participants to describe the orientation of force directions for several specified spinal discs.
We inquired how difficult they found these kinds of assessments and whether they believed directional encodings to be a useful addition to the tool.

After participants had used the tools during introduction and execution of tasks, we let them answer a questionnaire, where we asked how well they perceived correspondence with anatomical structures (\textbf{R1}), comparability of result data (\textbf{R2}), selection of points in time (\textbf{R3}) and the depiction of force directions (\textbf{R4}).
We associated each requirement with four to six statements the participants were to rate on a five-point Likert scale ($--$, $-$, $\circ$, $+$, $++$).
%
%We also asked which of our tool's data representation modes (e.g., perspective 3D or the simplified view) they would tend to use under which circumstances.
%
We concluded with a discussion about possible use cases for the proposed tool, to see, if the experts' opinions would match our designated application scenarios.
This was the only point in the evaluation where we distinguished between technical experts and medical practitioners, as their typical application domains would naturally differ.
\begin{table}[tb]
\caption{\label{tab:results}
    Results of the first and second task, comparing the former representation style (old) against our proposed exploration framework (new). Values are averaged over physicians and biomechanical simulation experts, as well as all participants (total).}
\begin{tabular*}{\linewidth}{l@{\extracolsep\fill}rrrr}
\hline
Task 1               & Mode & Total   & Physicians  & Experts   \\ \hline
Average error        & old  & 2.55N   & 5.12N       & 0.2N      \\
                     & new  & 0.19N   & 0.4N        & 0.03N     \\
Subjective certainty & old  & 52.8\%  & 36.1\%      & 69.4\%    \\
                     & new  & 97.2\%  & 94.4\%      & 100.0\%   \\
Time                 & old  & 6m 4s   & 6m 19s      & 5m 50s    \\
                     & new  & 3m 31s  & 3m 41s      & 3m 21s    \\ \hline
Task 2 \\ \hline 
Correct assessments  & old  & 66.7\%  & 38.9\%      & 94.4\%    \\
                     & new  & 100.0\% & 100.0\%     & 100.0\%   \\
Subjective certainty & old  & 45.8\%  & 16.7\%      & 75.0\%    \\
                     & new  & 95.8\%  & 91.7\%      & 100.0\%   \\
Time                 & old  & 3m 9s   & 2m 49s      & 3m 28s    \\
                     & new  & 1m 22s  & 1m 33s      & 1m 11s    \\ \hline
\end{tabular*}
\end{table}
\section{Results and Discussion}
\begin{figure}[tbp]
   \centering
   \includegraphics[width=0.95\linewidth]{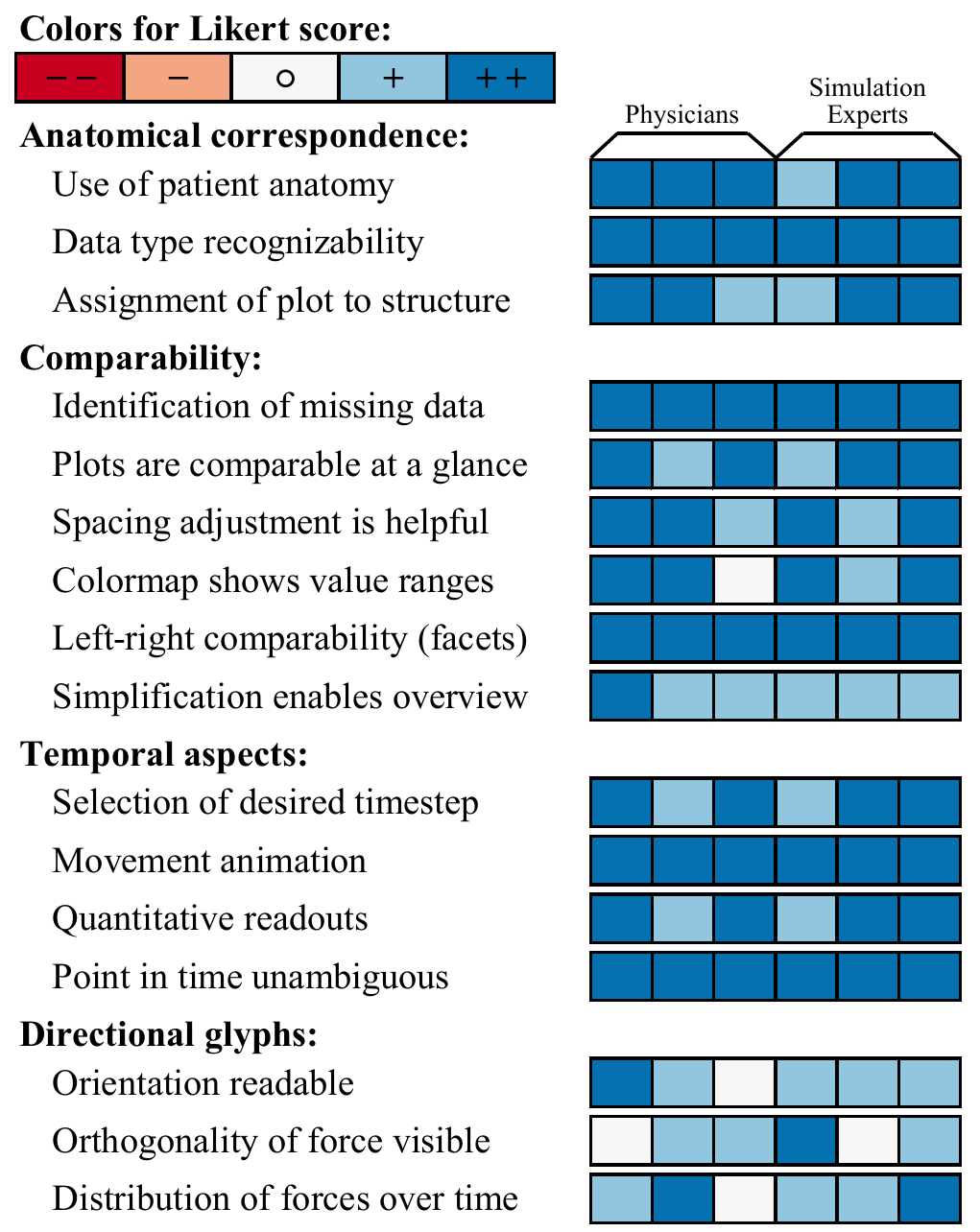}
   \caption{\label{fig:results}
    Results of the questionnaire with color-encoded Likert scores. Each box shows the answer of one participant.}
\end{figure}

All participants quickly understood most of the framework's functionality and were able to explore datasets without requiring assistance.
Unanimously, they deemed the tool a valuable asset they would like to use for analyzing spine simulations.
In the following, we describe some selected insights gained from discussion with the domain experts.
Additionally, we performed some measurements to compare the old representation style with our new tool and to validate participants' impressions.
These results do not represent a full quantitative study, but are meant to complement some findings of our qualitative interviews.
They are summarized in Table~\ref{tab:results}.
An overview of the questionnaire results is shown in Fig.~\ref{fig:results}.

%%%%%%%%%%%%%%%%%%%%%
\textbf{Anatomical correspondence.} Domain experts agreed that the visualization helps to foster a correspondence between simulation results and patient anatomy.
Especially the physicians pointed out that the clear connection between data and anatomy felt more intuitive to them than in the old representation.
They noted that this made them more inclined to actually use biomechanical simulation in practice.
We can affirm these impressions when looking at the average error (in Newton N) participants made when reading out values in the first task (see Table~\ref{tab:results}).
Even though both representations allow for quantitative readouts of results, errors occur when the wrong plot is selected.
This observation was even more evident during the second task: Physicians drew erroneous conclusions in more than half of the cases, when data was presented in the old format.
However, all six experts identified every single case correctly, when they were using our tool.
The most probable reason, why medical practicioners fared comparatively worse, is that our technical experts were already used to the old representation format and thus had an advantage.
The errors physicians made were almost all due to them inadvertently reading out the wrong plot or misinterpreting the scaling of an axis.
This disparity in accuracy (and also subjective certainty) could likely be overcome through training.
However, a direct connection between data and anatomy allowed physicians to immediately explore results, without having noticeable problems.
In the questionnaire, participants rated our representation of patient-specific vertebrae to be effective in facilitating intuitive assignments of data to structures (S($++$) = 5, S($+$) = 1).
They also found it easy to identify the type of displayed data, i.e., what the target structures are and which parameter is shown (S($++$) = 6).
This corresponds to our first requirement (\textbf{R1}).

%%%%%%%%%%%%%%%%%%%%%
\textbf{Comparisons.} All experts noted how our tool explicitly helped to compare result values of multiple spinal discs or facets.
They found the area charts to be comparable at a glance (S($++$) = 4, S($+$) = 2) and most rated the color map to be helpful to find data points with high loads (S($++$) = 4, S($+$) = 1, S($\circ$) = 1).
Every domain expert determined the two-sided plots to be highly useful when assessing facet parameters (S($++$) = 6).
Many stressed this point in particular already when performing the tasks.
Participants further confirmed that they could directly identify missing or faulty data (S($++$) = 6).
We can affirm this, as during the tasks one simulation expert and one clinician could not identify a missing facet data set when using the old representation, even when encouraged to do so.
However, all participants immediately pointed out this error when using the visualization.

%%%%%%%%%%%%%%%%%%%%%
\textbf{Animation.} The connection between movement and result values was also perceived well.
Participants said they found the real-time animation we display in a corner to be highly effective in this regard (S($++$) = 6).
They could quickly select a sought time step (S($++$) = 4, S($+$) = 2) and they felt it was obvious to understand which point in time was currently active (S($++$) = 6).

%%%%%%%%%%%%%%%%%%%%%
\textbf{Directional glyphs.} While overall responses were generally highly positive, we found participants to have most difficulties in the third task, when using the impact direction glyphs.
Some experts expressed a need to further familiarize themselves with the depiction, before they could accurately use it.
Still, most found the direction of acting forces on a spinal disc to be assessable (S($++$) = 1, S($+$) = 4, S($\circ$) = 1) and stated that they could determine the orientation of load distribution across several discs (S($++$) = 2, S($+$) = 2, S($\circ$) = 1).
We also observed that experts usually interpreted the direction correctly, however, many required 3D scene interaction to do so.
One physician mentioned that it would be advantageous if shear forces could be spotted more easily.
Note that at this point some features were not yet implemented, such as the isolines we now draw onto the spinal disc surfaces.
An aspect to take into account is that glyphs are different to methods currently used in practice.
In a real scenario, experts would therefore require more time to learn how a visualized force distribution appears in a physiological versus pathological case.
This means that simulation experts may already utilize such representations, for instance, to see if forces behave as expected and to detect errors.
For clinical use, it would be necessary to further explore what force direction characteristics different pathological cases exhibit, as compared to normal spines.
Then, it may be evaluated if such a depiction can be used to make better assessments and facilitate understanding of mechanical causes w.r.t. spine pathologies.

%%%%%%%%%%%%%%%%%%%%%
\textbf{Simplified view.} Experts mostly found the simplified depiction to provide a faster overview (S($++$) = 1, S($+$) = 5).
However, only some of them used this view during exploration.
We discussed probable reasons with the experts, who noted that it would be easier to understand data magnitudes when using the positional encoding, since the height differences of plots seemed more natural for them to compare as when they had to rely on color only.
Nonetheless, the simplified view has a definitive advantage regarding required screen space.
Even small depictions, as in Fig.~\ref{fig:simplified}, remain readable.
Therefore, we came to the consensus that this representation would be particularly suited to compare multiple data sets.
Future evaluations should thus target this use case explicitly.

%%%%%%%%%%%%%%%%%%%%%
\textbf{Application scenarios.} When discussing possible application scenarios for the proposed visualizations, all physicians pointed out the possibility of treatment evaluation through comparison of data collected pre- and post-surgery.
They could also imagine the simulation of different treatment options, e.g., contrasting different possibilities for implants and analyzing resulting loads on critical structures.
Some proposed supplementing traditional diagnostics by examining simulated loads on a patient's spine and the possibility of supporting physician-patient communication through intuitive data representations.
Biomechanical simulation experts deemed the tool to be highly promising for evaluating result data, identifying computation errors and visually comparing effects of different simulation parameters.
They also suggested the use in interdisciplinary communication of simulation results.
These scenarios are reflective of what we intended the framework to be used for.
\section{Improvements}
Based on the evaluation we revised and modified some of the employed encodings.
These improvements were subsequently discussed with the domain experts we collaborated with for the initial requirement analysis.

\textbf{Directional glyphs.}
As participants appeared to require 3D scene interaction in order to read the directional glyphs, we adapted the representation.
When the glyphs are rendered, we reduce the vertebrae geometry to its silhouette projection, in order to visually guide focus towards the directional glyphs.
We debated adding some form of encoding showing deviation of force directions from a spinal centerline, similar to the ideas of Klemm et al.~\cite{klemm_2013_VMV, klemm_2014_tvcg, klemm_2015_ivapp}.
The problem we encountered is that a centerline would, again, be somewhat dependent on the viewing angle.
From a frontal (coronal) direction, deviations towards the left and right could be shown.
However, deviations towards the front or back would require another perspective.
Instead, we propose projecting isolines in parallel direction to the force plane onto the spinal disc model.
These lines make vertical forces immediately discernible from shear forces (cf. Fig.~\ref{fig:glyphs},~\ref{fig:glyphs_shear}).
Additionally they can be displayed even if the spine is in its original state.
We believe that with some experience a domain expert would already be able to hint at force directions by interpreting the isolines, without having to artificially increase the distance between vertebra, in order to see the full directional glyphs.
We deliberately use arrows, discs and isolines as redundant mappings of spatial direction.
We found their combination to reduce the risk of information loss and make the glyphs comprehensible from a static point of view, as demonstrated in Fig.~\ref{fig:glyphs_shear}.
One simulation expert stated that it might be useful if the amount of jitter in force direction could be visualized, as this would be a valuable indicator of simulation stability.
To emphasize the glyph movements, we render the arrows' trajectory in form of a traced surface geometry, which fades away at selection of later time-points.
A jittering force direction can be immediately spotted, as it results in a wide surface.
Impact directions, which are stable over time, produce only a small surface trace.

The additions were positively received by the experts, who found the difference between shear forces and vertical impacts to be readable more clearly.
They also agreed that the isolines are a helpful extension to understand the directions even in a ``compressed'' model of the spine, where the full glyphs cannot be shown.

\textbf{Simplified view.}
Realizing the potential of the simplified view mode for ensemble and multi-set visualizations, we emphasized this use-case by removing detail from the scene when the user intends to compare more than two data sets.
Individual values are probably less important in this scenario, so we disabled the output of quantified values at the selected time-step per default.
Also, the vertebral geometry is de-emphasized by using a flat projection.
Additionally, we found subtle changes in the color gradient to be hard to discern in these downscaled views, which is why we employed a discretized color scheme.
The value range of the color map and the discretization level is configurable by the user.
The experts acknowledged that the discretization helps to better distinguish values in small scale depictions.
They affirmed that this also enables to spot where certain thresholds are exceeded, which is helpful to simulation experts and clinicians alike.
\section{Conclusion and Future Work}
In this design study, we presented a framework for visual exploration of human spine biomechanics.
The visualizations target intuitive representations of time-dependent parameters, which are simulated using patient-specific anatomy.
We aim to support simulation researchers in understanding computed data and make clinically relevant properties accessible to spinal surgeons.
We proposed a combination of interactive charts, glyphs and a simplified representation to make simulation results explorable in depth, as well as to enhance data overview and comparison.

An evaluation with six domain experts showed that our tool has potential to complement research in biomechanical spine simulation and may provide a way to introduce simulations into medical practice.
%
%We found an increase in accuracy and a distinct reduction in time expenditure when experts used our tools to evaluate data sets as compared to traditional methods of standard plotting.
%
We suggested novel glyph-based depictions of spatial force distributions that were not apparent from the data before.
The visualizations solely rely on data that is already used in the simulation pipeline: the raw output data, the segmented vertebrae models, and (optional) movement patterns.
This makes our methods generalizable regarding additional spinal parameters and means they could also be used within different spine simulation workflows.

We believe that some of our insights could be transferred to other areas as well.
In particular, this applies to the concept of connecting data with anatomy in the field of medical visualization, which appears to be a strong motivator for clinicians to adapt new encodings, as they can more intuitively understand them.
Embedding of abstract data representations in a context familiar to the user allows for a structural interpretation.
This is a research direction that we believe still holds potential, not only in the medical domain.
Making sense of large amounts of data, acquired through sensors or simulation, can be facilitated through a combination of techniques from information and scientific visualization, resulting in hybrid representations, as they are shown in this work.
A challenge is the combined display of data with differing dimensions, e.g., 2D graphs in a 3D scene. A possible solution shown here is to fix 3D viewpoints and to manipulate or deform geometry to enforce data alignment.
To improve readability of such visualizations, we found the following principles to be particularly useful: colormaps for data overview and abstraction, focus and context to adapt to specific exploration tasks, and use of symmetry to foster intuitive comparison if two sides are naturally given.
Also, the use of embedded glyphs to visualize impact directions could be transferred to more types of mechanical or biomechanical simulations, where researchers would benefit from getting a better understanding of force vector orientations.

In the future, we would like to evaluate an expansion of our methods w.r.t. to full spines, as compared to the cervical spine examples we used in this work.
Further, we would like to incorporate additional information into the animation window.
Computing and displaying important structural relations, angles and degrees of freedom could make multiple data sets quantifiably comparable regarding motion data.
For instance, this would allow to contrast resulting forces on spinal discs in dependency of a defined bending angle.
Also, we would like to explore additional methods for a combined overview of spine geometry and simulation results, in order to replace the potentially flawed depiction of stacked graphs.
As mentioned before, we currently simulate patient movements through external forces as a proof-of-concept.
While our results should scale well with movement data from visible light scans, which is currently being integrated into the pipeline, we would like to test patient-specific motion data with the proposed tools once this becomes an option.
This way, we aim to advance personalized medicine, by simulating accurate patient spine geometry in combination with individual movement patterns and enabling physicians to intuitively explore the resulting data.

\acknowledgments{
This work was partially funded by the Carl Zeiss Foundation and the Federal Ministry for Economic Affairs and Energy of Germany.}

\bibliographystyle{abbrv-doi.bst}

\bibliography{template.bib}

\begin{thebibliography}{10}

\bibitem{al2018automatic}
I.~Al-Dhamari, S.~Bauer, and D.~Paulus.
\newblock Automatic multi-modal cervical spine image atlas segmentation.
\newblock In A.~Maier, T.~M. Deserno, H.~Handels, K.~H. Maier-Hein, C.~Palm,
  and T.~Tolxdorff, eds., {\em Bildverarbeitung f{\"u}r die Medizin}, pp.
  303--308. Springer, 2018. doi: {{%
10\hspace{.1pt}\discretionary{.}{%
}{.}\hspace{.4pt}1007\discretionary{/}{%
}{/}978\discretionary{%
}{-}{-}3\discretionary{%
}{-}{-}662\discretionary{%
}{-}{-}56537\discretionary{%
}{-}{-}7\_80}}


\bibitem{andersson1999epidemiological}
G.~B. Andersson.
\newblock Epidemiological features of chronic low-back pain.
\newblock {\em Lancet}, 354(9178):581--585, 1999.

\bibitem{ay20133d}
M.~Ay, T.~Kubat, C.~Delilbasi, B.~Ekici, H.~E. Yuzbasioglu, and
  S.~Hartomacioglu.
\newblock {3D} {Bio-CAD} modeling of human mandible and fabrication by
  rapid-prototyping technology.
\newblock {\em Usak University Journal of Material Sciences}, 2:135--145, 2013.
  doi: {{%
10\hspace{.1pt}\discretionary{.}{%
}{.}\hspace{.4pt}12748\discretionary{/}{%
}{/}uujms\hspace{.1pt}\discretionary{.}{%
}{.}\hspace{.4pt}201324255}}


\bibitem{bauer2016basics}
S.~Bauer.
\newblock Basics of multibody systems: Presented by practical simulation
  examples of spine models.
\newblock In R.~Lopez-Ruiz, ed., {\em Numerical Simulation - From Brain Imaging
  to Turbulent Flows}. IntechOpen, London, 2016. doi: {{%
10\hspace{.1pt}\discretionary{.}{%
}{.}\hspace{.4pt}5772\discretionary{/}{%
}{/}62864}}


\bibitem{bauer2013biomechanical}
S.~Bauer and U.~Buchholz.
\newblock Biomechanical effects of spinal fusion to adjacent vertebral
  segments.
\newblock In {\em 2013 European Modelling Symposium}, pp. 158--163. IEEE, 2013.
  doi: {{%
10\hspace{.1pt}\discretionary{.}{%
}{.}\hspace{.4pt}1109\discretionary{/}{%
}{/}EMS\hspace{.1pt}\discretionary{.}{%
}{.}\hspace{.4pt}2013\hspace{.1pt}\discretionary{.}{%
}{.}\hspace{.4pt}28}}


\bibitem{bauer2015analysis}
S.~Bauer and D.~Paulus.
\newblock Analysis of the biomechanical effects of spinal fusion to adjacent
  vertebral segments of the lumbar spine using multi body simulation.
\newblock {\em International Journal of Simulation-Systems, Science and
  Technology}, 15, 2015.

\bibitem{bauer2015does}
S.~Bauer and D.~Paulus.
\newblock How does the intervertebral discs parameter variation affect the
  biomechanical behavior of spinal structures? results of a detailed study of
  multibody simulation sensitivity.
\newblock {\em International Journal of Engineering and Applied Sciences},
  2(9):37--42, 2015. doi: {{%
10\hspace{.1pt}\discretionary{.}{%
}{.}\hspace{.4pt}1515\discretionary{/}{%
}{/}cdbme\discretionary{%
}{-}{-}2015\discretionary{%
}{-}{-}0092}}


\bibitem{bauer2017computational}
S.~Bauer and D.~Paulus.
\newblock Computational simulation as an innovative approach in personalized
  medicine.
\newblock In J.~Bettany-Saltikov and S.~Schreiber, eds., {\em Computational
  Simulation as an Innovative Approach in Personalized Medicine, Innovations in
  Spinal Deformities and Postural Disorders}. IntechOpen, London, 2017. doi:
  {{%
10\hspace{.1pt}\discretionary{.}{%
}{.}\hspace{.4pt}5772\discretionary{/}{%
}{/}intechopen\hspace{.1pt}\discretionary{.}{%
}{.}\hspace{.4pt}68835}}


\bibitem{bauer2014quantification}
S.~Bauer, C.~Wasserhess, and D.~Paulus.
\newblock Quantification of loads on the lumbar spine of children with
  different body weight — a comparative study with the help of computer
  modelling.
\newblock {\em Biomedical Engineering}, 59:910--1027, 2014. doi: {{%
10\hspace{.1pt}\discretionary{.}{%
}{.}\hspace{.4pt}1515\discretionary{/}{%
}{/}bmt\discretionary{%
}{-}{-}2014\discretionary{%
}{-}{-}5012}}


\bibitem{bennett2007aesthetics}
C.~Bennett, J.~Ryall, L.~Spalteholz, and A.~Gooch.
\newblock The aesthetics of graph visualization.
\newblock {\em Computational Aesthetics}, 2007:57--64, 2007.

\bibitem{Borgo2013}
R.~Borgo, J.~Kehrer, D.~H. Chung, E.~Maguire, and {et al.}
\newblock {Glyph-based visualization: Foundations, design guidelines,
  techniques and applications}.
\newblock {\em Eurographics}, pp. 39--63, 2013.

\bibitem{cuellar2017text}
J.~M. Cu{\'e}llar and T.~H. Lanman.
\newblock “text neck”: an epidemic of the modern era of cell phones?
\newblock {\em The Spine Journal}, 17(6):901--902, 2017. doi: {{%
10\hspace{.1pt}\discretionary{.}{%
}{.}\hspace{.4pt}1016\discretionary{/}{%
}{/}j\hspace{.1pt}\discretionary{.}{%
}{.}\hspace{.4pt}spinee\hspace{.1pt}\discretionary{.}{%
}{.}\hspace{.4pt}2017\hspace{.1pt}\discretionary{.}{%
}{.}\hspace{.4pt}03\hspace{.1pt}\discretionary{.}{%
}{.}\hspace{.4pt}009}}


\bibitem{delp2007opensim}
S.~L. Delp, F.~C. Anderson, A.~S. Arnold, P.~Loan, A.~Habib, C.~T. John,
  E.~Guendelman, and D.~G. Thelen.
\newblock Opensim: open-source software to create and analyze dynamic
  simulations of movement.
\newblock {\em IEEE Transactions on Biomedical Engineering}, 54(11):1940--1950,
  2007.

\bibitem{Demir2014}
I.~Demir, C.~Dick, and R.~Westermann.
\newblock Multi-charts for comparative {3D} ensemble visualization.
\newblock {\em IEEE Trans Vis Comput Graph}, 20(12):2694--2703, 2014.

\bibitem{Dick:2011:DistanceVisualization}
C.~Dick, R.~Burgkart, and R.~Westermann.
\newblock Distance visualization for interactive {3D} implant planning.
\newblock {\em IEEE Trans Vis Comput Graph}, 17(12):2173--2182, 2011.

\bibitem{Dick:2009:HipJointReplacementPlanning}
C.~Dick, J.~Georgii, R.~Burgkart, and R.~Westermann.
\newblock A {3D} simulation system for hip joint replacement planning.
\newblock In {\em Proceedings of World Congress on Medical Physics and
  Biomedical Engineering 2009}, vol. 25/IV of {\em IFMBE Proceedings}, pp.
  363--366, 2009.

\bibitem{Dick:2009:StressTensorFields}
C.~Dick, J.~Georgii, R.~Burgkart, and R.~Westermann.
\newblock Stress tensor field visualization for implant planning in
  orthopedics.
\newblock {\em IEEE Trans Vis Comput Graph}, 15(6):1399--1406, 2009.

\bibitem{Doleisch2007}
H.~Doleisch.
\newblock Simvis: Interactive visual analysis of large and time-dependent {3D}
  simulation data.
\newblock In {\em Winter Simulation Conference}, pp. 712--20, 2007.

\bibitem{Doleisch2003}
H.~Doleisch, M.~Gasser, and H.~Hauser.
\newblock Interactive feature specification for focus+context visualization of
  complex simulation data.
\newblock In {\em Proc. of IEEE/EG VisSym}, pp. 239--48, 2003.

\bibitem{Doleisch2004}
H.~Doleisch, M.~Mayer, M.~Gasser, R.~Wanker, and H.~Hauser.
\newblock Case study: Visual analysis of complex, time-dependent simulation
  results of a diesel exhaust system.
\newblock In {\em Proc. of IEEE/EG VisSym}, pp. 91--6, 2004.

\bibitem{eulzer2019temporal}
P.~Eulzer, S.~Engelhardt, N.~Lichtenberg, R.~De~Simone, and K.~Lawonn.
\newblock Temporal views of flattened mitral valve geometries.
\newblock {\em IEEE Trans Vis Comput Graph}, 26(1):971--980, 2019.

\bibitem{fishwick1992simpack}
P.~A. Fishwick.
\newblock Simpack: getting started with simulation programming in c and c++.
\newblock In {\em Winter Simulation Conference}, pp. 154--162, 1992.

\bibitem{freyd1984force}
J.~Freyd and B.~Tversky.
\newblock Force of symmetry in form perception.
\newblock {\em The American Journal of Psychology}, pp. 109--126, 1984.

\bibitem{gatchel2015continuing}
R.~J. Gatchel.
\newblock The continuing and growing epidemic of chronic low back pain.
\newblock {\em Healthcare}, 3(3):838--845, 2015. doi: {{%
10\hspace{.1pt}\discretionary{.}{%
}{.}\hspace{.4pt}3390\discretionary{/}{%
}{/}healthcare3030838}}


\bibitem{ghiselli2004adjacent}
G.~Ghiselli, J.~C. Wang, N.~N. Bhatia, W.~K. Hsu, and E.~G. Dawson.
\newblock Adjacent segment degeneration in the lumbar spine.
\newblock {\em JBJS}, 86(7):1497--1503, 2004.

\bibitem{grob2007association}
D.~Grob, H.~Frauenfelder, and A.~Mannion.
\newblock The association between cervical spine curvature and neck pain.
\newblock {\em European Spine Journal}, 16(5):669--678, 2007. doi: {{%
10\hspace{.1pt}\discretionary{.}{%
}{.}\hspace{.4pt}1007\discretionary{/}{%
}{/}s00586\discretionary{%
}{-}{-}006\discretionary{%
}{-}{-}0254\discretionary{%
}{-}{-}1}}


\bibitem{who2003burden}
W.~S. Group.
\newblock The burden of musculoskeletal conditions at the start of the new
  millennium.
\newblock Technical Report 919, World Health Organization, 2003.

\bibitem{hermann2015}
M.~Hermann, A.~C. Schunke, T.~Schultz, and R.~Klein.
\newblock Accurate interactive visualization of large deformations and
  variability in biomedical image ensembles.
\newblock {\em IEEE Trans Vis Comput Graph}, 22(1):708--717, 2016.

\bibitem{heuch2013body}
I.~Heuch, I.~Heuch, K.~Hagen, and J.-A. Zwart.
\newblock Body mass index as a risk factor for developing chronic low back
  pain.
\newblock {\em Spine}, 38(2):133--139, 2013. doi: {{%
10\hspace{.1pt}\discretionary{.}{%
}{.}\hspace{.4pt}1097\discretionary{/}{%
}{/}BRS\hspace{.1pt}\discretionary{.}{%
}{.}\hspace{.4pt}0b013e3182647af2}}


\bibitem{kettler2006validity}
A.~Kettler, F.~Rohlmann, C.~Neidlinger-Wilke, K.~Werner, L.~Claes, and H.-J.
  Wilke.
\newblock Validity and interobserver agreement of a new radiographic grading
  system for intervertebral disc degeneration: Part ii. cervical spine.
\newblock {\em European Spine Journal}, 15(6):732--741, 2006. doi: {{%
10\hspace{.1pt}\discretionary{.}{%
}{.}\hspace{.4pt}1007\discretionary{/}{%
}{/}s00586\discretionary{%
}{-}{-}005\discretionary{%
}{-}{-}1037\discretionary{%
}{-}{-}9}}


\bibitem{kim2017}
K.~Kim, J.~V. Carlis, and D.~F. Keefe.
\newblock Comparison techniques utilized in spatial {3D} and {4D} data
  visualizations: A survey and future directions.
\newblock {\em Computers \& Graphics}, 67:138--147, 2017.

\bibitem{klemm_2015_ivapp}
P.~Klemm, S.~Gla{\ss}er, K.~Lawonn, M.~Rak, H.~V{\"{o}}lzke, K.~Hegenscheid,
  and B.~Preim.
\newblock Interactive visual analysis of lumbar back pain.
\newblock In {\em Proc. of the 6th International Conference on Information
  Visualization Theory and Applications}, pp. 85--92. Berlin, 2015.

\bibitem{klemm_2013_VMV}
P.~Klemm, K.~Lawonn, M.~Rak, B.~Preim, K.~T{\"{o}}nnies, K.~Hegenscheid,
  H.~V{\"{o}}lzke, and S.~Oeltze.
\newblock Visualization and analysis of lumbar spine canal variability in
  cohort study data.
\newblock In {Michael Bronstein, Jean Favre, and Kai Hormann}, ed., {\em VMV
  2013 - Vision, Modeling, Visualization}, pp. 121--128. Lugano, 2013.

\bibitem{klemm_2014_tvcg}
P.~Klemm, S.~Oeltze-Jafra, K.~Lawonn, K.~Hegenscheid, H.~V{\"{o}}lzke, and
  B.~Preim.
\newblock Interactive visual analysis of image-centric cohort study data.
\newblock {\em IEEE Trans Vis Comput Graph}, pp. 1673--1682, 2014. doi: {{%
10\hspace{.1pt}\discretionary{.}{%
}{.}\hspace{.4pt}1109\discretionary{/}{%
}{/}TVCG\hspace{.1pt}\discretionary{.}{%
}{.}\hspace{.4pt}2014\hspace{.1pt}\discretionary{.}{%
}{.}\hspace{.4pt}2346591}}


\bibitem{Konyha2012}
Z.~Konyha, A.~Le{\v{z}}, K.~Matkovi{\'c}, M.~Jelovi{\'c}, and H.~Hauser.
\newblock Interactive visual analysis of families of curves using data
  aggregation and derivation.
\newblock In {\em Proc. of the Conference on Knowledge Management and Knowledge
  Technologies}, p.~24, 2012.

\bibitem{krekel2006interactive}
P.~R. Krekel, C.~P. Botha, E.~R. Valstar, P.~W. de~Bruin, P.~M. Rozing, and
  F.~H. Post.
\newblock Interactive simulation and comparative visualisation of the
  bone-determined range of motion of the human shoulder.
\newblock In {\em SimVis}, pp. 275--288, 2006.

\bibitem{Krekel2010VisualAO}
P.~R. Krekel, E.~R. Valstar, J.~H.~D. Groot, F.~H. Post, R.~G. H.~H. Nelissen,
  and C.~P. Botha.
\newblock Visual analysis of multi-joint kinematic data.
\newblock {\em Comput Graph Forum}, 29:1123--1132, 2010.

\bibitem{kumar2001correlation}
M.~Kumar, A.~Baklanov, and D.~Chopin.
\newblock Correlation between sagittal plane changes and adjacent segment
  degeneration following lumbar spine fusion.
\newblock {\em European Spine Journal}, 10(4):314--319, 2001. doi: {{%
10\hspace{.1pt}\discretionary{.}{%
}{.}\hspace{.4pt}1007\discretionary{/}{%
}{/}s005860000}}


\bibitem{Lawonn_2014_CGFb}
K.~Lawonn, R.~Gasteiger, and B.~Preim.
\newblock Adaptive surface visualization of vessels with animated blood flow.
\newblock {\em Comput Graph Forum}, 33(8):16--27, 2014.

\bibitem{Lawonn_2015_TVCG}
K.~Lawonn, S.~Gla{\ss}er, A.~Vilanova, B.~Preim, and T.~Isenberg.
\newblock Occlusion-free blood flow animation with wall thickness
  visualization.
\newblock {\em IEEE Trans Vis Comput Graph}, 22 (1)(1):728--737, 2015.

\bibitem{Lawonn_2015_Feature}
K.~Lawonn and B.~Preim.
\newblock {\em Feature Lines for Illustrating Medical Surface Models:
  Mathematical Background and Survey}, chap. Visualization in Medicine in Life
  Sciences III, pp. 93--132.
\newblock Springer Verlag, 2016.

\bibitem{doi:10.1111/cgf.13306}
K.~Lawonn, N.~N. Smit, K.~Bühler, and B.~Preim.
\newblock A survey on multimodal medical data visualization.
\newblock {\em Comput Graph Forum}, 37(1):413--438, 2018. doi: {{%
10\hspace{.1pt}\discretionary{.}{%
}{.}\hspace{.4pt}1111\discretionary{/}{%
}{/}cgf\hspace{.1pt}\discretionary{.}{%
}{.}\hspace{.4pt}13306}}


\bibitem{Lawonn_2018_CGF}
K.~Lawonn, I.~Viola, B.~Preim, and T.~Isenberg.
\newblock A survey of surface-based illustrative rendering for visualization.
\newblock {\em Comput Graph Forum}, 37(6):205--234, 2018.

\bibitem{liu2018somewhere}
Y.~Liu and J.~Heer.
\newblock Somewhere over the rainbow: An empirical assessment of quantitative
  colormaps.
\newblock In {\em Proceedings of the 2018 CHI Conference on Human Factors in
  Computing Systems}, pp. 1--12, 2018.

\bibitem{Meuschke_2017_TVCG}
M.~Meuschke, S.~Vo{\ss}, O.~Beuing, B.~Preim, and K.~Lawonn.
\newblock Combined visualization of vessel deformation and hemodynamics in
  cerebral aneurysms.
\newblock {\em IEEE Trans Vis Comput Graph}, 23(1):761--770, 2017.

\bibitem{Meuschke:2017:GCS:3128397.3128407}
M.~Meuschke, S.~Vo{\ss}, O.~Beuing, B.~Preim, and K.~Lawonn.
\newblock Glyph-based comparative stress tensor visualization in cerebral
  aneurysms.
\newblock {\em Comput Graph Forum}, 36(3):99--108, June 2017. doi: {{%
10\hspace{.1pt}\discretionary{.}{%
}{.}\hspace{.4pt}1111\discretionary{/}{%
}{/}cgf\hspace{.1pt}\discretionary{.}{%
}{.}\hspace{.4pt}13171}}


\bibitem{mikkonen2013association}
P.~H. Mikkonen, J.~Laitinen, J.~Remes, T.~Tammelin, S.~Taimela, K.~Kaikkonen,
  P.~Zitting, R.~Korpelainen, and J.~Karppinen.
\newblock Association between overweight and low back pain: a population-based
  prospective cohort study of adolescents.
\newblock {\em Spine}, 38(12):1026--1033, 2013. doi: {{%
10\hspace{.1pt}\discretionary{.}{%
}{.}\hspace{.4pt}1093\discretionary{/}{%
}{/}aje\discretionary{/}{%
}{/}kwp356}}


\bibitem{panjabi1990clinical}
M.~Panjabi and A.~White.
\newblock Clinical biomechanics of the spine.
\newblock {\em Kinematics of the Spine}, pp. 85--127, 1990.

\bibitem{Potter2009}
K.~Potter, A.~Wilson, P.-T. Bremer, D.~Williams, C.~Doutriaux, V.~Pascucci, and
  C.~R. Johnson.
\newblock Ensemble-vis: A framework for the statistical visualization of
  ensemble data.
\newblock In {\em IEEE International Conference on Data Mining Workshops}, pp.
  233--40, 2009.

\bibitem{raidou2018}
R.~G. Raidou, O.~Casares-Magaz, A.~Amirkhanov, V.~Moiseenko, L.~P. Muren, J.~P.
  Einck, A.~Vilanova, and M.~E. Gr{\"o}ller.
\newblock Bladder runner: Visual analytics for the exploration of rt-induced
  bladder toxicity in a cohort study.
\newblock In {\em Comput Graph Forum}, vol.~37, pp. 205--216, 2018.

\bibitem{rajaee2012spinal}
S.~S. Rajaee, H.~W. Bae, L.~E. Kanim, and R.~B. Delamarter.
\newblock Spinal fusion in the united states: Analysis of trends from 1998 to
  2008.
\newblock {\em Spine}, 37(1):67--76, 2012. doi: {{%
10\hspace{.1pt}\discretionary{.}{%
}{.}\hspace{.4pt}1097\discretionary{/}{%
}{/}BRS\hspace{.1pt}\discretionary{.}{%
}{.}\hspace{.4pt}0b013e31820cccfb}}


\bibitem{rengier20103d}
F.~Rengier, A.~Mehndiratta, H.~Von Tengg-Kobligk, C.~M. Zechmann,
  R.~Unterhinninghofen, H.-U. Kauczor, and F.~L. Giesel.
\newblock {3D} printing based on imaging data: review of medical applications.
\newblock {\em International Journal of Computer Assisted Radiology and
  Surgery}, 5(4):335--341, 2010. doi: {{%
10\hspace{.1pt}\discretionary{.}{%
}{.}\hspace{.4pt}1007\discretionary{/}{%
}{/}s11548\discretionary{%
}{-}{-}010\discretionary{%
}{-}{-}0476\discretionary{%
}{-}{-}x 2010}}


\bibitem{rohlmann1999internal}
A.~Rohlmann, J.~Calisse, G.~Bergmann, and U.~Weber.
\newblock Internal spinal fixator stiffness has only a minor influence on
  stresses in the adjacent discs.
\newblock {\em Spine}, 24(12):1192--1196, 1999.

\bibitem{Ropinski2011}
T.~Ropinski, S.~Oeltze, and B.~Preim.
\newblock {Survey of glyph-based visualization techniques for spatial
  multivariate medical data}.
\newblock In {\em {Computers \& Graphics}}, vol.~35, pp. 392--401, 2011.

\bibitem{roussouly2005classification}
P.~Roussouly, S.~Gollogly, E.~Berthonnaud, and J.~Dimnet.
\newblock Classification of the normal variation in the sagittal alignment of
  the human lumbar spine and pelvis in the standing position.
\newblock {\em Spine}, 30(3):346--353, 2005. doi: {{%
10\hspace{.1pt}\discretionary{.}{%
}{.}\hspace{.4pt}1097\discretionary{/}{%
}{/}01\hspace{.1pt}\discretionary{.}{%
}{.}\hspace{.4pt}brs\hspace{.1pt}\discretionary{.}{%
}{.}\hspace{.4pt}0000152379\hspace{.1pt}\discretionary{.}{%
}{.}\hspace{.4pt}54463\hspace{.1pt}\discretionary{.}{%
}{.}\hspace{.4pt}65}}


\bibitem{solteszova2020memento}
V.~Solteszova, N.~N. Smit, S.~Stoppel, R.~Gr{\"u}ner, and S.~Bruckner.
\newblock Memento: Localized time-warping for spatio-temporal selection.
\newblock In {\em Computer Graphics Forum}, vol.~39, pp. 231--243, 2020.

\bibitem{sun2005bio}
W.~Sun, B.~Starly, J.~Nam, and A.~Darling.
\newblock Bio-cad modeling and its applications in computer-aided tissue
  engineering.
\newblock {\em Computer-Aided Design}, 37(11):1097--1114, 2005.

\bibitem{van2008perceptual}
F.~van Ham and B.~Rogowitz.
\newblock Perceptual organization in user-generated graph layouts.
\newblock {\em IEEE Trans Vis Comput Graph}, 14(6):1333--1339, 2008.

\bibitem{ware2018measuring}
C.~Ware, T.~L. Turton, R.~Bujack, F.~Samsel, P.~Shrivastava, and D.~H. Rogers.
\newblock Measuring and modeling the feature detection threshold functions of
  colormaps.
\newblock {\em IEEE Trans Vis Comput Graph}, 25(9):2777--2790, 2018.

\bibitem{Weissenbock2018}
J.~Weissenb{\"o}ck, B.~Fr{\"o}hler, E.~Gr{\"o}ller, J.~Kastner, and C.~Heinzl.
\newblock Dynamic volume lines: Visual comparison of {3D} volumes through
  space-filling curves.
\newblock {\em IEEE Trans Vis Comput Graph}, 25(1):1040--49, 2018.

\bibitem{wilke1994universal}
H.-J. Wilke, L.~Claes, H.~Schmitt, and S.~Wolf.
\newblock A universal spine tester for in vitro experiments with muscle force
  simulation.
\newblock {\em European Spine Journal}, 3(2):91--97, 1994.

\bibitem{wilke2006validity}
H.-J. Wilke, F.~Rohlmann, C.~Neidlinger-Wilke, K.~Werner, L.~Claes, and
  A.~Kettler.
\newblock Validity and interobserver agreement of a new radiographic grading
  system for intervertebral disc degeneration: Part i. lumbar spine.
\newblock {\em European Spine Journal}, 15(6):720--730, 2006. doi: {{%
10\hspace{.1pt}\discretionary{.}{%
}{.}\hspace{.4pt}1007\discretionary{/}{%
}{/}s00586\discretionary{%
}{-}{-}005\discretionary{%
}{-}{-}1029\discretionary{%
}{-}{-}9}}


\bibitem{zhang2015glyph}
C.~Zhang, T.~Schultz, K.~Lawonn, E.~Eisemann, and A.~Vilanova.
\newblock Glyph-based comparative visualization for diffusion tensor fields.
\newblock {\em IEEE Trans Vis Comput Graph}, 22(1):797--806, 2015.

\end{thebibliography}
\end{document}